# Distributional bias compromises leave-one-out cross-validation

George I. Austin[1,2], Itsik Pe'er[2,3], Tal Korem[2,4,†]

**Author affiliations**

[1] Department of Biomedical Informatics, Columbia University Irving Medical Center, New York, NY, USA

[2] Program for Mathematical Genomics, Department of Systems Biology, Columbia University Irving Medical Center, New York, NY, USA

[3] Department of Computer Science, Columbia University, New York, NY, USA

[4] Department of Obstetrics and Gynecology, Columbia University Irving Medical Center, New York, NY, USA

[†] Corresponding author: tal.korem@columbia.edu

# Abstract

Cross-validation is a common method for estimating the predictive performance of machine learning models. In a data-scarce regime, where one typically wishes to maximize the number of instances used for training the model, an approach called "leave-one-out cross-validation" is often used. In this design, a separate model is built for predicting each data instance after training on all other instances. Since this results in a single test instance available per model trained, predictions are aggregated across the entire dataset to calculate common performance metrics such as the area under the receiver operating characteristic or $R^2$ scores. In this work, we demonstrate that this approach creates a negative correlation between the average label of each training fold and the label of its corresponding test instance, a phenomenon that we term distributional bias. As machine learning models tend to regress to the mean of their training data, this distributional bias tends to negatively impact performance evaluation and hyperparameter optimization. We show that this effect generalizes to leave-P-out cross-validation and persists across a wide range of modeling and evaluation approaches, and that it can lead to a bias against stronger regularization. To address this, we propose a generalizable rebalanced cross-validation approach that corrects for distributional bias for both classification and regression. We demonstrate that our approach improves cross-validation performance evaluation in synthetic simulations, across machine learning benchmarks, and in several published leave-one-out analyses.

# Introduction

One of the most important challenges of devising any machine learning model, whether it is as simple as a linear model over structured data or as complex as a large language model applied to free text, is to robustly evaluate its performance[1–9]. A common strategy for such evaluations is "cross-validation"[10–13]. In cross-validation, the data is divided into parts ("folds", e.g., a tenth of the data), which are then each used as a held-out test set for a model trained on the rest of the dataset. The predictions made by the model on each held-out test set are then compared to the hidden labels, either separately for each fold or in aggregate across folds via metrics such as the mean squared error, $R^2$, area under the receiver operator characteristic curve[14] (auROC) or area under the precision-recall curve[15] (auPR; **Fig. S1)**. Optimal approaches for cross-validation is a heavily studied topic[11,12,16–27]. Most approaches have well-documented and common drawbacks, such as the potential for overly optimistic performance estimates[16–18], limited stability[13] or high computational costs[20,28,29].

Training and evaluating models in data-scarce scenarios, in which the acquisition of every data instance is challenging or expensive, is difficult. In datasets that are small to begin with, reducing the data further by holding out a fold as a test set makes it significantly harder for the model to infer the underlying signal, reducing performance. To keep nearly all samples available for training each model, a common variation of cross validation, known as "leave-one-out cross-validation" (LOOCV), is typically used[11,21,25–27,30,31]. LOOCV treats every individual data instance as an independent test set, thus maximizing the number of samples used for training each model. For small datasets, the high computational cost of LOOCV is less of an issue, and it is often recommended for several key advantages, including high consistency, stability across estimators[22,23,30], and a low propensity for a bias towards overly optimistic performance estimations[12,31,32]. LOOCV can also be generalized for keeping any number of samples as a test fold, known as "leave-P-out cross-validation"[23,33,34] (LPOCV), where P is the number of samples left-out per test fold.

In this work, we demonstrate that standard implementations of LPOCV introduce distributional bias – a negative correlation, across folds, between the average label of the corresponding training and test sets (briefly observed before[24]) – and show that it negatively affects performance evaluation. We demonstrate this phenomenon via a simple predictor which takes advantage of this bias to obtain perfect scores on any LOOCV task. In practice, however, due to the tendency of machine learning models to regress to the mean of the training set labels, this bias tends to decrease common performance evaluation metrics, such as auROC, auPR, or $R^2$, particularly when these are evaluated across folds. We show that this impacts not just performance evaluation, but also hyperparameter optimization, with distributional bias favoring the selection of models with weaker regularization and impacting nested cross-validation. To address this issue, we propose a generalizable correction for LOOCV, called "Rebalanced LOOCV", (RLOOCV). We demonstrate that this correction consistently increases measured machine learning performances under multiple machine learning benchmarks, including published LOOCV-based evaluation of models predicting preterm birth, chronic fatigue syndrome, and adverse events for immune checkpoint inhibitor treatment.

# Results

### Distributional bias leaks information on test labels in LOOCV and LPOCV

When using LOOCV of a dataset with $N$ samples with average label $\mu$, every removal of a sample $y_j$ as a test set changes the mean of the training set labels, $\mu_{-j} = \frac{\mu N - y_j}{N-1}$. Across cross-validation folds, the decrease in $\mu_{-j}$ is proportional to $y_j$, creating a perfect negative correlation (r = -1) between the means of the training set labels and the labels of the test sets. In this scenario, information about the test label "leaks" into the training set: the held-out labels can be uniquely inferred by simply observing the training label average. In the case of binary classification ($y_j \in \{0,1\}$; **Fig. 1a**), the only two possible values for $\mu_{-j}$ are $\frac{\mu N-1}{N-1}$ and $\frac{\mu N}{N-1}$. Therefore, when running standard LOOCV, the $\mu N$ folds in which the held-out $y_j$ is 1 will have a training label mean of $\frac{\mu N-1}{N-1}$, and the $N(1-\mu)$ folds in which the held-out $y_j$ is 0 will have a training label mean of $\frac{\mu N}{N-1}$. It follows that in any LOOCV applied for the purpose of binary classification, there is a shift of $\frac{1}{N-1}$ in the label average of the training dataset that is inversely correlated to the held-out label $y_j$. To demonstrate this leakage, we simulated random data and showed that a dummy model that only predicts the negative mean of the training set's labels achieves perfect auROC and auPR of 1 (**Figs. 1b, S2a; Methods**). This is regardless of the underlying data, and in a manner completely independent from the test data and labels. We focus most of our investigation below on classification tasks, and end by demonstrating generalization to regression.

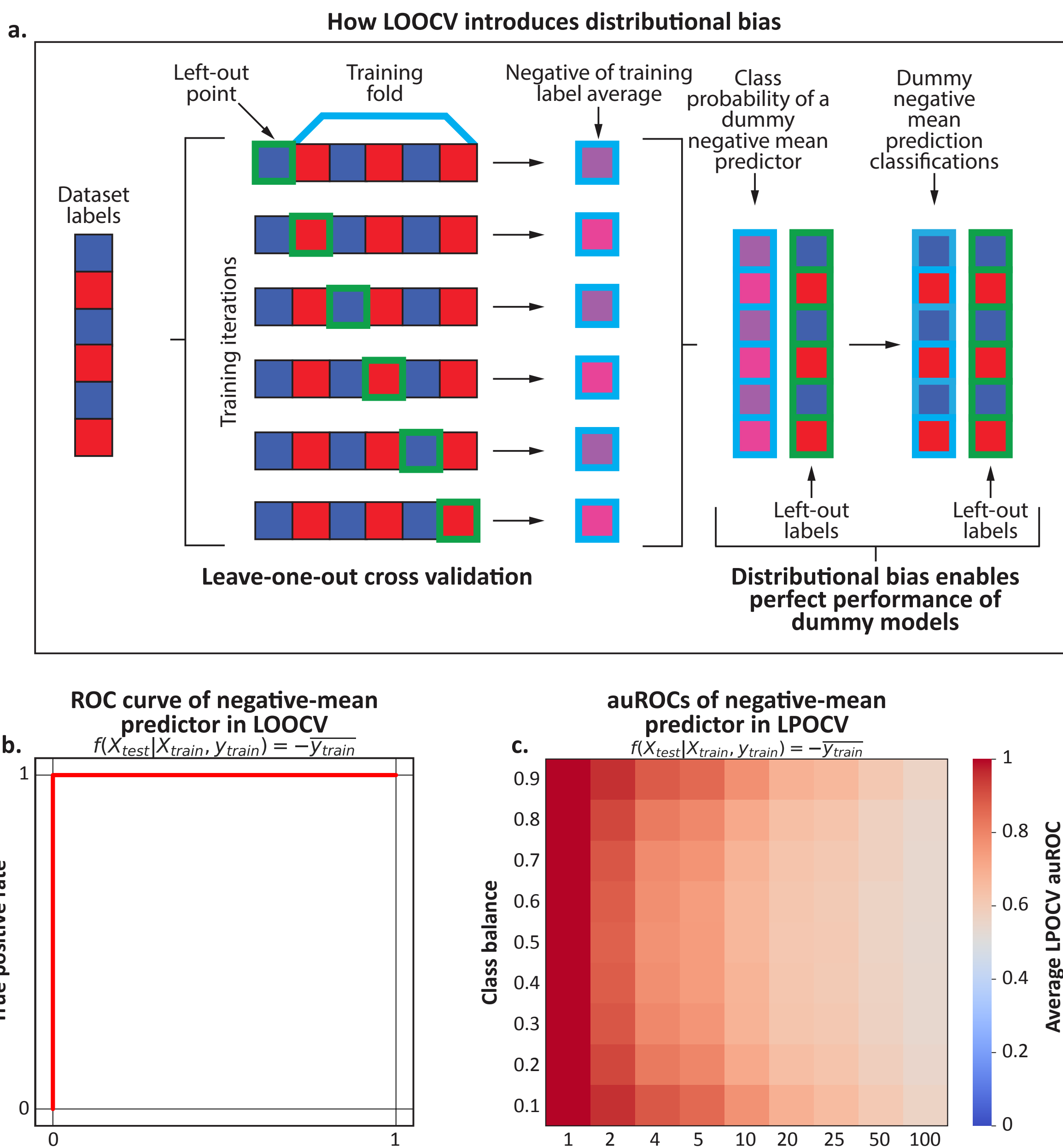

**Figure 1 | Distributional bias leaks the test set label in leave-one-out cross-validation. a,** An illustration of how distributional bias manifests in LOOCV. When a held-out data instance belongs to either class, the class average of the remaining dataset shifts by $\frac{1}{N-1}$ in the other direction. As a result, a dummy predictor that returns the negative of the average training class label would produce predictions that are perfectly correlated with the actual labels. **b,** Receiver operator characteristic (ROC) curve for this dummy negative-mean predictor. The auROC is 1 under any scenario regardless of the underlying data. **c,** A heatmap showing the average auROC of the same dummy negative-mean predictor under different class balances and P-left-out schema on randomly generated labels, with resulting auROCs consistently over the expected null auROC of 0.5. Each cell in the heatmap shows the results of 100 independent simulations.

Next, we sought to evaluate whether this effect generalizes to more general LPOCV scenarios. To do so, we simulated random datasets of 250-300 samples while varying the class balance (from 10% to 90%) and the number of samples left out per fold (from 1 to 100; **Methods**). Evaluating the same dummy model on this data, we observed that the impact of distributional bias decreases as the fold size increases, starting with auROCs of 1.00 at N=1 and decreasing to an average auROC of 0.55 (range of 0.54-0.57) at N=100 (**Fig. 1c**). However, even at N=100, the dummy predictor still obtained significantly higher auROCs than a random guess (one-sided one-sample t-test p<0.001), indicating that distributional bias is still an issue even in this setting. We further observed an interaction between the class balance, held-out P size, and the impact of distributional bias, with higher impact on more extreme class balances. For example, the dummy model for P=4 had mean±std auROCs of 0.88±0.03 and 0.88±0.03 for class balances of 10% and 90%, respectively, while it had a mean±std auROCs of 0.76±0.04 for the same P at a class balance of 50% (**Figs. 1c, S2b**). Complementarily, a different dummy predictor that outputs the mean of the training set labels obtains the worst possible auROC and auPR in this scenario (**Fig. S2c,d**). Overall, our results indicate a strong information leakage through distributional bias with LOOCV, and that the effect is present even when the held-out test set has more than one sample. Furthermore, we demonstrate this effect to be more pronounced when dealing with significant class imbalances.

**Distributional bias leads to lower performance evaluation of common machine learning models**

After observing how distributional bias can be used by an adversarial dummy model specifically built to take advantage of this phenomenon, we sought to investigate its impact on the performance evaluation of standard machine learning models. We therefore generated similar simulations as above, with all features drawn i.i.d from uniform distributions ∈ [0,1]. Because, by construction, all features and labels are random, any fair assessment of a machine learning model on this data should yield an average auROC of 0.5. However, when assessing the performance of an $L^2$-regularized logistic regression model, a random forest model, and a K-Nearest Neighbor model (K=5) in LOOCV, we instead found auROCs significantly lower than 0.5 across all class balances (one-sample t-test $p < 0.01$ for all vs. 0.5, **Figs. 2a, S3**). Distributional bias manifests in this way because the models we used tend to predict values close to the class balance, a phenomenon known as regression to the mean (**Fig. S4)**. The effect of distributional bias was more pronounced on logistic regression models (mean±std auROC of 0.23±0.13 across all simulations; **Fig. 2a**) than on random forests (0.48±0.08; **Fig. S3a**) and 5-Nearest Neighbor (0.48±0.04; **Fig. S3b**), in line with stronger regression to the mean observed for the former (**Fig. S4**). Overall, our results demonstrate that distributional bias tends to decrease the performance of common machine learning models when evaluated using LOOCV.

**Exact stratification corrects distributional bias**

As distributional bias results from minor shifts in the class balance of the training set, it would not be present in any evaluation approach that ensures an identical class balance across either the training or test sets of different folds. To demonstrate this, we first implemented a stratified LPOCV approach, in which we made best effort, given P and the class balance, to maintain an identical class balance across all folds. When exact stratification was possible, this approach completely resolved distributional bias and led to a fair performance evaluation (**Fig. 2c**). For example, a stratified leave-2-out CV with a class balance of 50% had a mean±std auROC of 0.50±0.05, and a stratified leave-5-out scheme had auROCs of 0.50±0.06 for 20%, 40%, 60%, and 80% (aggregated; p=0.59 via one-sample t-test vs. 0.5; **Fig. 2b**). However, when the class balance cannot be precisely stratified due to a particular combination of P, class balance, and dataset size (e.g., leave-2-out with a class balance of 10%), small shifts in label means will still occur, leading to an under-evaluation of performance. For example, in the leave-5-out scheme and class balances of 10%, 30%, 50%, 70%, and 90%, we observe that all auROCs were below 0.45 (**Fig. 2b**). These results demonstrate that stratification can correct for distributional bias in LPOCV only when the choice of P facilitates exact stratification.

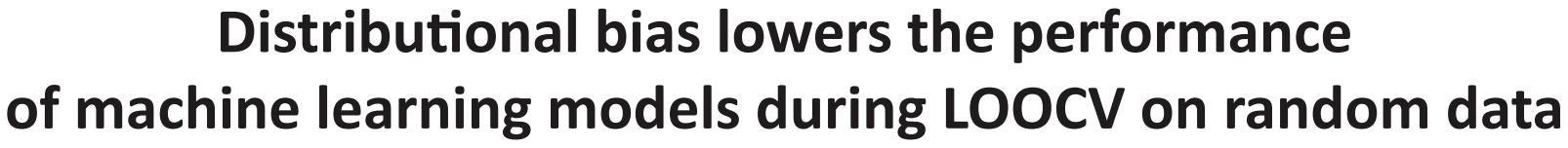

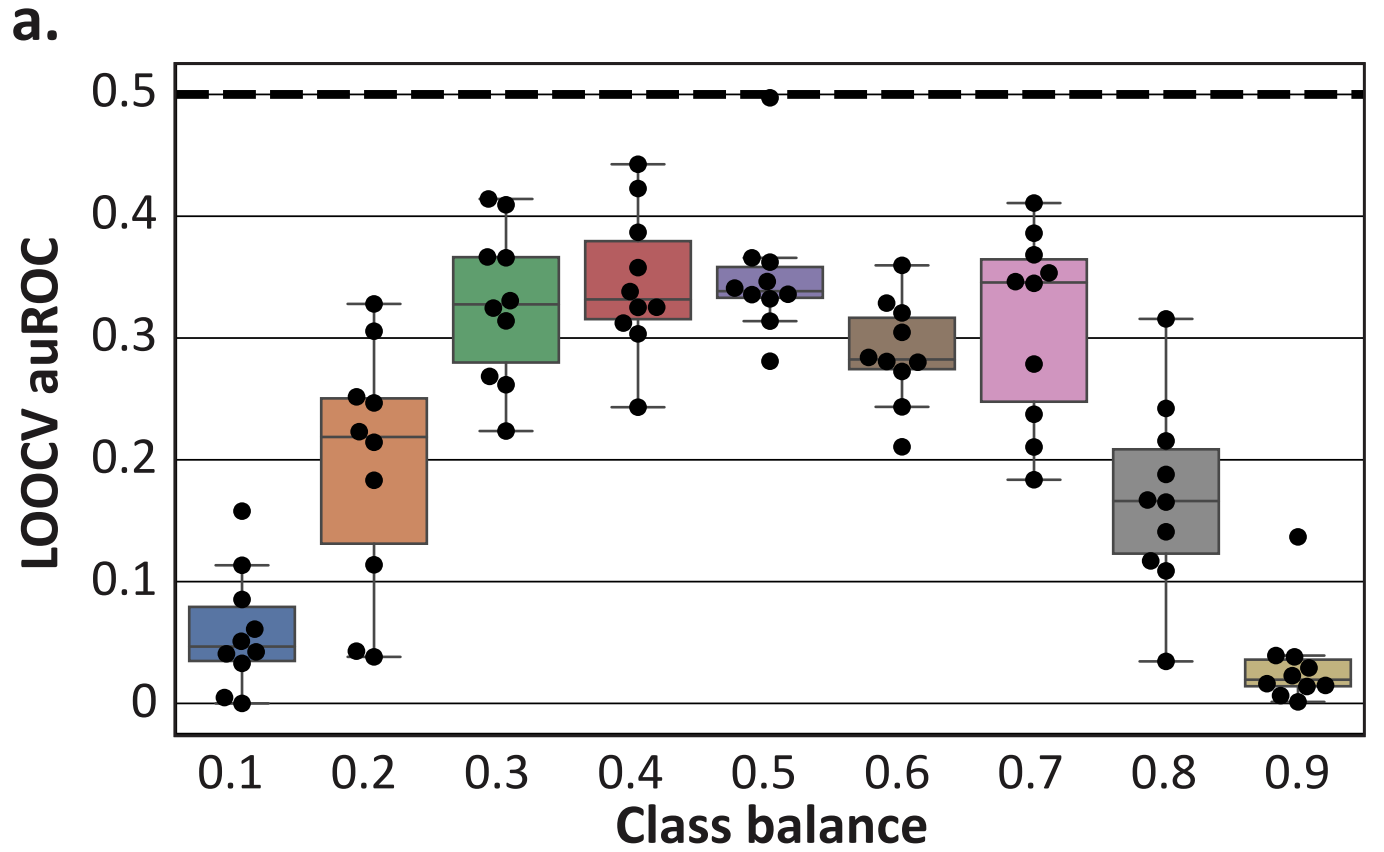

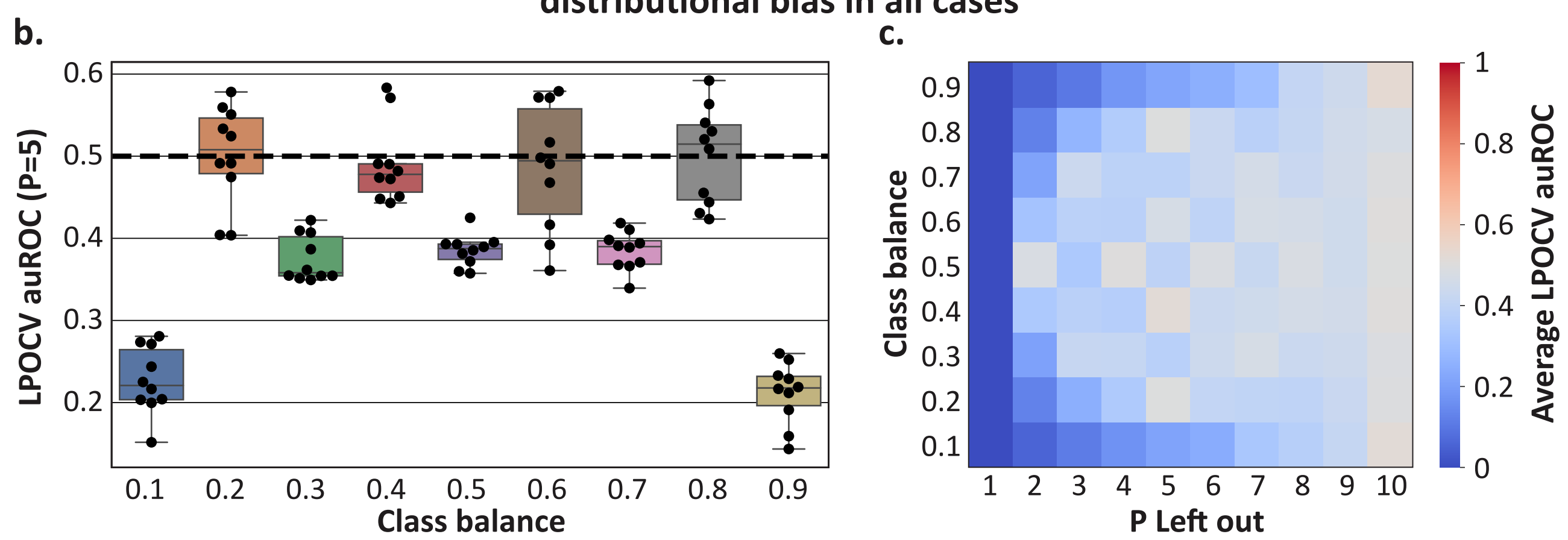

**Figure 2 | Distributional bias produces results worse than a random guess on random data.** All plots pertain to LOOCV and LPOCV analyses of logistic regression models on randomly generated data and labels. The auROC in this setting should be 0.5 in any fair evaluation. In **a,b,** one point corresponds to one simulated dataset; in **c**, one cell corresponds to 10 simulations. **a**, Boxplots of auROCs for a standard LOOCV implementation across different underlying class balances. Resulting auROCs are consistently less than 0.5 (aggregated $p<0.001$ via a single 1-sample t-test). **b**, Boxplots of auROCs on stratified leave-five-out cross-validation across different class balances. When the class balances can be precisely captured with five samples (e.g., class balance of 0.2), the distribution of resulting auROCs has a mean which is not significantly different from 0.5 (one-sample t test vs. 0.5 p=0.59). Otherwise, under-evaluation of performance is evident (e.g., for class balance of 0.1). **c**, Heatmap of average auROCs on stratified LPOCV for Ps ranging from 1 to 10 and for different class balances. Results demonstrate that the effect of distributional bias, observed as auROCs below 0.5, is smaller the closer the stratification enabled by P and the class balance is to optimal stratification. Box, IQR; line, median; whiskers, nearest point to 1.5*IQR.

### A rebalanced leave-one-out CV for distributional-bias correction

As we demonstrated that stratification cannot always address the effects of distributional bias, we developed a generalizable approach to correct it. We propose a Rebalanced Leave-One-Out Cross Validation ("RLOOCV"), in which we subsample the training dataset to remove a randomly selected data instance of the opposite label as each held-out test instance (**Fig. 3a**). This ensures that all training sets have the same class balance (even if that balance is slightly different than the class balance across the entire dataset). We show that RLOOCV can also be generalized to LPOCV (**Supplementary Note 1**). We note that some variations of LOOCV that are not sensitive to distributional bias have been proposed (for purposes other than addressing distributional bias), such as a stratified leave-pair-out CV with resampling in which auROCs are calculated separately for each held-out test set[21]. Another approach could be to up-sample the training set with a sample of the held-out class to achieve a similar balance, avoiding loss of samples for training. However, up-sampling and data augmentation involve domain specific knowledge and assumptions[35–37] and are less generalizable.

To test RLOOCV, we first checked how it evaluates the dummy negative-mean predictor on random data. We found that as RLOOCV maintains a constant training-set label mean across all folds, and therefore resolves any potential distributional shifts, the auROC of this adversarial predictor was 0.5 across all class balances (**Fig. 3b**). Of note, this is different from a stratified leave-2-out CV, which can only address a class balance of 0.5 (**Fig. 2c**). We observed a similar effect when using RLOOCV to evaluate a logistic regression model on random simulated data across different class balances. While LOOCV evaluated the same model with auROCs consistently lower than 0.5 (**Fig. 2a**), auROCs evaluated with RLOOCV are not significantly different from 0.5 ($p = 0.84$ via a single aggregated one-sample t-test, **Fig. 3c**), as expected for random data. Overall, our synthetic simulations demonstrate that the impact of distributional bias can be corrected by strategic subsampling within the training set, which does not pose any risk of information leakage or of any erroneous inflation of predictive results.

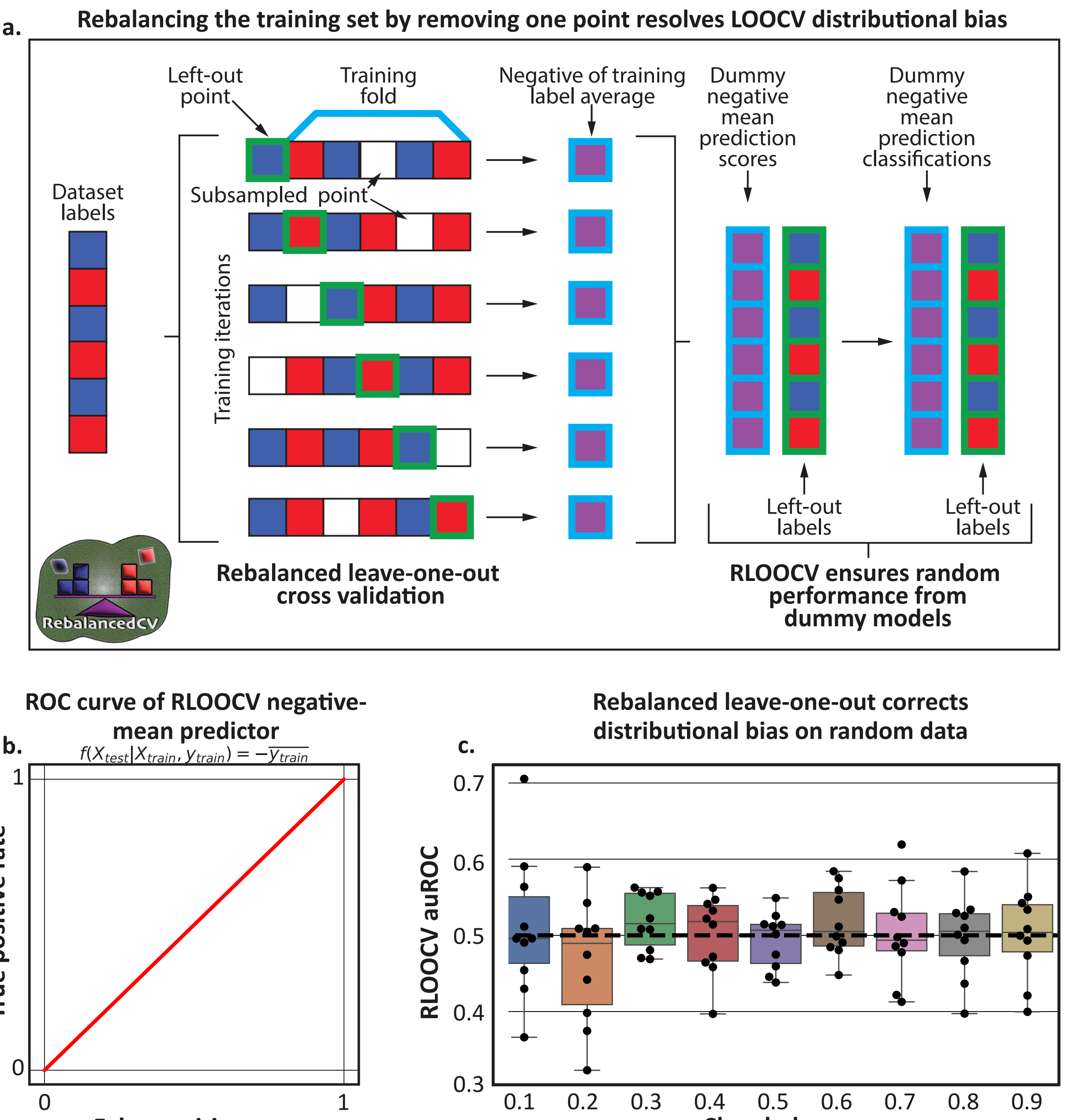

**Figure 3 | Rebalancing training data through subsampling avoids distributional bias. a**, An illustration of our proposed rebalanced LOOCV (RLOOCV) for classification. For each test instance (or fold), we remove from the training set a data instance with the opposite label such that the training set's label mean is constant across all folds. This can be accomplished by randomly removing a single training instance with a label opposite that of the test instance. **b,** ROC curve of the negative-mean predictor (similar to **Fig. 1b**) evaluated via RLOOCV, which resulted in an auROC of 0.50 (the expected result for an evaluation of a dummy predictor). **c,** Boxplots (Box, IQR; line, median; whiskers, nearest point to

1.5*IQR) of auROCs of a logistic regression model trained on randomly generated data, similar to **Fig. 2a**, but evaluated with RLOOCV. The resulting auROCs are not consistently higher or lower than the expected 0.5 ($p$=0.84 via a single aggregated one-sample t-test).

**Rebalancing the training set improves models evaluated on real datasets**

After evaluating RLOOCV on simulated data, we sought to evaluate its impact on a broad range of realistic datasets. To this end, we analyzed 49 datasets from the UC Irvine Machine Learning Repository[38] (UCIMLR; **Methods**). We found a consistent improvement in performance when using RLOOCV instead of LOOCV to train and evaluate $L^2$-regularized logistic regression models (median [IQR] auROCs of 0.74 [0.60-1.00] for LOOCV vs 0.81 [0.69-1.00] for RLOOCV; two-sided Wilcoxon signed-rank $p=3.9\times10^{-5}$; **Fig. 4a**). When a model performed well on a dataset (auROCs at or close to 1), we generally found similar performance in both LOOCV and RLOOCV, as such models would have minimal regression to the mean. Conversely, on harder tasks (lower auROCs), we see stronger improvements with RLOOCV. Specifically, we found multiple datasets in which models trained in LOOCV obtained an auROC of 0, while no model trained with RLOOCV obtained an auROC below 0.4. This is likely due to the fact that these more difficult training tasks will have a greater propensity to regress to the training-set label mean, which will give distributional bias a greater influence.

To further emphasize this effect, we used the same modeling approach, but provided the models with only the first two principal components of the features. In this setting we expect a stronger regression to the mean, and indeed, we found that models trained in LOOCV often had much worse performances (median [IQR] auROC of 0.73 [0.06-0.97]), while the impact on models trained in RLOOCV was much smaller (auROC of 0.77 [0.54-0.97]; two-sided Wilcoxon signed-rank $p=1.7\times10^{-7}$ vs. LOOCV; **Fig. 4a**). These results further demonstrate the impact of distributional bias on real datasets, and confirm that RLOOCV can improve analyses in a wide variety of machine learning settings.

Next, we evaluated the effect of RLOOCV on bespoke models trained using LOOCV by other investigators on real biological datasets. To this end, we selected a few cases with available processed data and either code or clear methods. First, we analyzed classifiers of preterm birth from vaginal microbiome samples[39], using processed data published elsewhere[40]. Using nested 5-fold CV for tuning and evaluation of logistic regression models, we found that while using LOOCV yielded a median auROC of 0.692 (median of 10 bootstrap runs), the median auROC increased to 0.697 with RLOOCV (**Fig. 4b**). We also evaluated prediction of immune-related adverse events (irAEs) in melanoma patients undergoing immune checkpoint blockade using bulk T cell receptor diversity and activated CD4 memory T cell abundances[41]. While repeating the original analysis obtained a median auROC of 0.817 using logistic regression evaluated using LOOCV, we again found a slight increase to a median auROC of 0.833 using RLOOCV (**Fig. 4c**).

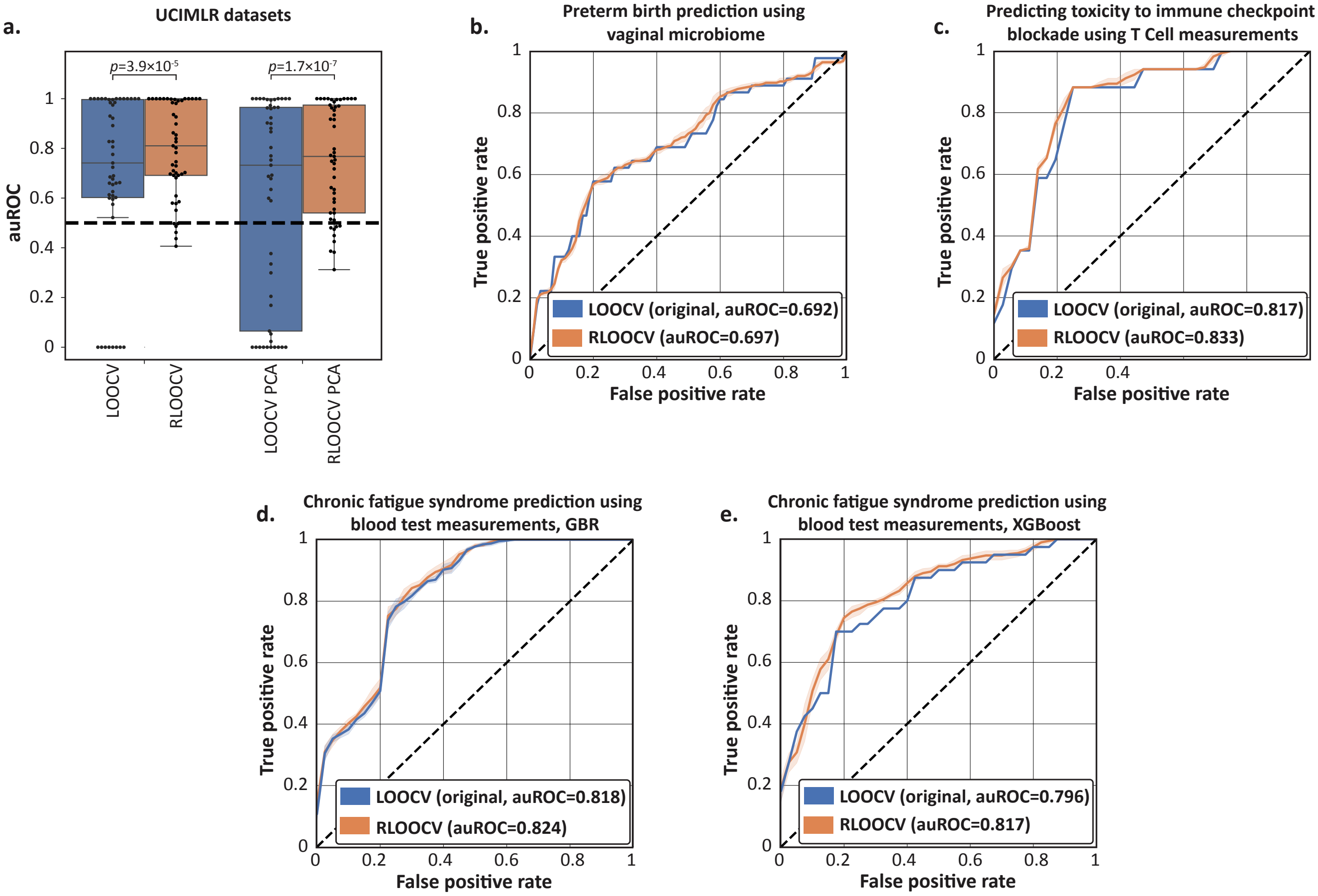

**Figure 4 | Correcting distributional bias with RLOOCV improves performance evaluation of published predictive models.** (**a**) auROCs (y-axis) of $L^2$-regularized logistic regression models trained in cross-validation on multiple classification benchmarks from UCIMLR (**Methods**). "PCA" denotes results of models that were provided only with the first two principal components, which are less expressive and have a stronger tendency to regress to the mean. (**b-e**) ROC curves comparing the performance of published models evaluated with LOOCV with the same models evaluated using our rebalancing approach (RLOOCV) over 10 bootstrap runs. Tasks include predicting preterm birth from vaginal microbiome samples using logistic regression[39,40] (**b**); predicting complications from immune checkpoint blockade therapy using T cell measurements[41], also using logistic regression (**c**); and predicting chronic fatigue syndrome from standard blood test measurements[41,42] using gradient boosted regression (**d**) and XGBoost (**e**). Across all cases, we observed a small but consistent improvement from RLOOCV (Fisher's multiple comparison of DeLong tests $p$=0.015 across all four evaluations). Shaded areas represent 95% confidence intervals.

Next, we sought to examine RLOOCV evaluation of more expressive models, such as gradient boosted regression[43,44]. We therefore replicated an analysis which classified chronic fatigue syndrome from 34 standard blood test measurements[42]. The original analysis trained a gradient boosted regression model, which, when replicated and evaluated with LOOCV, had a median auROC of 0.818. Using RLOOCV, this evaluation increased to 0.824 (**Fig. 4d**). The original analysis also evaluated XGBoost models, which yielded a median auROC of 0.796. Here we saw a slightly larger improvement with RLOOCV, to a median auROC of 0.817 (**Fig. 4e**). Altogether, across four different analyses with diverse data types, we observe consistent improvements using RLOOCV implementation (Fisher's multiple comparison of DeLong[45] tests $p$=0.015 across all four evaluations). We therefore propose RLOOCV as a low-risk alternative to LOOCV that successfully addresses distributional bias in predictive settings.

**Distributional bias affects hyperparameter optimization**

We demonstrated consistent improvements in performance evaluation obtained by correcting distributional biases. While these improvements are small, it could be argued that they are also technical, in that they do not stem from improved predictive capacity of the model itself and therefore would not have a substantial impact on downstream implementations. However, cross-fold evaluation metrics are often used not just for performance evaluation, but also for hyperparameter optimization and model selection. To evaluate the potential effects of distributional bias on this optimization, we next focused on regularization, which is often key to obtaining robust model performance.

We first examined our simulations with random data (**Fig. 2**). Comparing logistic regression models with different regularization strengths using LOOCV and different class balances, we found a stronger impact of distributional bias on models with higher regularization, with auROC=0 in all of the high regularization settings, compared to a mean±std auROC of 0.48±0.07 at the lowest regularization (**Fig. 5a**). This is because less expressive models with heavier regularization are more likely to predict values close to the label mean, and therefore are more likely to see decreased results due to this distributional shift. Switching to RLOOCV, however, removes the effects of distributional bias. Using the same model and regularization parameters, we observed mean±std auROCs of 0.50±0.05 across a wide range of regularization strength (Fisher's combined probability test for 13 one-sample t tests vs 0.5 across model parameters p=0.23; **Fig. 5b).**

Next, we hypothesized that as distributional bias has greater performance reduction on models with higher regularization (**Fig. 5a,b**), it would result in selection of weaker regularization parameters in hyperparameter optimization. To evaluate this hypothesis, we considered the dataset we analyzed in **Fig. 4c**, in which irAEs were predicted from circulating T-cells characteristics[41], and evaluated a range of regularization strengths for logistic regression models. We found that auROCs evaluated with RLOOCV were significantly higher than those evaluated with LOOCV across all settings (Wilcoxon signed-rank p=0.0012), with the best auROCs obtained with RLOOCV (auROC of 0.845 compared to 0.817 with LOOCV; **Fig. 5c**). Interestingly, the regularization level that yielded the best auROCs was different between LOOCV and RLOOCV, with distributional bias causing the best predictor selected with LOOCV to be less regularized than the best predictor selected with RLOOCV (optimal regularization strength of $10^{-6}$-$10^{-2}$ vs. $10^{2}$-$10^{5}$, respectively, **Fig. 5c**). Consistent with our simulations, the impact of distributional bias is particularly evident in the high-regularization settings, in which distributional bias carries relatively higher importance. In those scenarios, which could often be relevant for inference, we see LOOCV-evaluated auROCs of 0, while the auROCs evaluated with RLOOCV can have optimal performance (**Fig. 5c**). Overall, these results demonstrate that distributional bias would cause hyperparameter optimization performed using non-balanced cross-validation to select suboptimal regularization parameters.

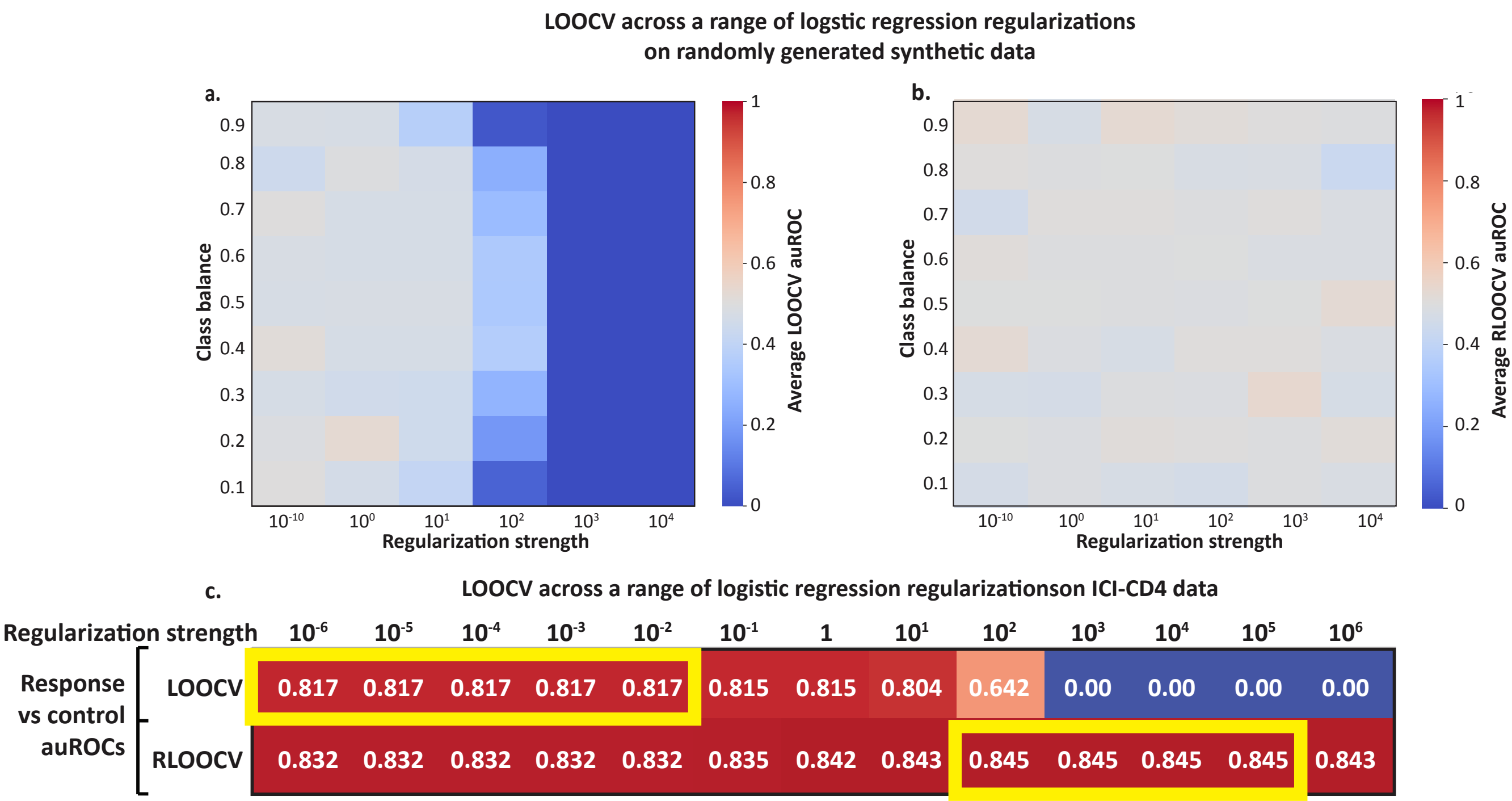

**Figure 5 | Distributional bias and LOOCV favor weaker regularization. a,b,** Heatmaps pertain to analyses of logistic regression models on randomly generated data and labels, where the auROC should be 0.5 in any fair evaluation. **a**, Average auROCs evaluated with LOOCV across varying regularization strength and class balances, which are consistently less than 0.5 ($p$<0.001 via one-sample t-test across all values). **b,** Same heatmap as in (**a**), but with RLOOCV. Resulting auROCs are not consistently higher or lower than 0.5 (Fisher's combined probability test across six independent one-sample t tests vs. 0.5 p=0.23). **c,** Heatmap showing the auROC obtained by logistic regression models classifying patients who experienced complications from immune checkpoint blockade therapy using T cell measurements (**Methods**). Different rows correspond to evaluation using LOOCV and RLOOCV, while different columns correspond to different regularization strengths. The optimal performance in each setup was obtained by RLOOCV. Additionally, the optimal performance under LOOCV was obtained with weaker regularization compared to evaluation with RLOOCV, suggesting that distributional bias can cause models tuned via LOOCV to be less regularized.

Finally, we evaluated whether distributional bias can also impact hyperparameter optimization under a nested cross-validation setting. We first evaluated the same *in silico* simulations used in **Figs. 2-3**, in which models should perform randomly. We used $L^2$-regularized logistic regression models, and tuned the regularization strength with either: (1) outer LOOCV and inner LOOCV (known as "Honest LOOCV"[31]); (2) outer and inner RLOOCV ("Nested RLOOCV"); (3) outer LOOCV with inner 5-fold cross validation ("5-fold LOOCV"); and (4) outer RLOOCV with inner 5-fold cross-validation ("5-fold RLOOCV"). Across three class balances, outer LOOCV, either with "Honest LOOCV" or with inner 5-fold, performed consistently worse than random (one-sample t test vs. 0.5 $p<0.05$ for all comparisons; **Fig. S5a**). When the outer cross-validation was performed with RLOOCV, with either nested 5-fold CV or nested RLOOCV, performance was on par with that of a random guess (one-sample t test vs. 0.5 $p\geq0.1$ in all cases, **Fig. S5a**). The same set of evaluation benchmarks also demonstrated an improvement with RLOOCV when predicting ICI response using the data published by Lozano et al.[41] (analyzed in **Fig. 4c**), with models trained in RLOOCV obtaining auROCs of 0.835 and 0.838, compared to auROCs of 0.819 and 0.815 when applying Honest LOOCV and 5-fold LOOCV (**Fig. S5b**). Similarly, nested RLOOCV outperformed Honest LOOCV on UCIMLR benchmark (two-sided Wilcoxon signed-rank $p=1.4\times10^{-5}$ and $p=1.0\times10^{-7}$, **Fig. S5c**). As before, providing both models with only the first two principal components of the data yielded a stronger reduction in performance for models trained using nested LOOCV compared to those trained with nested RLOOCV, demonstrating that evaluation with LOOCV will favor more complex models. These results further highlight the impact of distributional bias for hyperparameter optimization, and the bias it creates towards selecting more complex models, demonstrating that it applies even in nested cross-validation.

**Distributional bias impacts regression and is alleviated by RLOOCV**

As mentioned above, under LOOCV, a perfect negative correlation ($r=-1$) emerges between the means of training set labels and those of the test-set labels, also impacting regression (**Fig. 6a,b**). However, while in classification tasks there is always a sample that can be removed from each training set to exactly offset distributional bias (given at least one sample in each class), this is not guaranteed for regression. Nevertheless, one can find a sample whose removal from the training set would shift its average label to be as close as possible to the average label of the entire dataset (visualized by the horizontal line in **Figs. 6b**). We note that rebalancing the training set such that its average label would shift beyond the global average would risk overinflating performance estimates. We have therefore implemented a version of RLOOCV for regression (**Fig. S6**), which selects the optimal data instance that would shift the average training set label as close to, but not past, the dataset label average, such that training sets whose corresponding held-out instance is greater than this value will always have a training average at or below it, and vice-versa. If, for a particular held-out label, no match exists whose subsampling would shift the remaining training dataset's mean towards, but not past the overall average, we subsample no additional data instances, and train on the full N-1 samples. RLOOCV substantially reduces the negative correlation between training label average and held-out test instance (**Fig. 6b**).

Next, we sought to evaluate how distributional bias manifests during *in silico* simulations in which the known ground truth is random performance, and evaluated the performance of RLOOCV. To this end, we generated random data and labels as in **Figs. 2-3** (**Methods**), with all features and labels drawn from normal distributions, and evaluated $L^2$-regularized linear regression models with LOOCV and RLOOCV. We found that, as with classification, models evaluated with LOOCV performed significantly worse than random (median [IQR] $R^2$ of -0.016 [-0.022-0.010]; one-sample t test vs. 0 $p=7.1\times10^{-8}$; **Fig. 6c**). Evaluations with RLOOCV, while still worse than random (p=0.0017) were consistently closer to the known ground-truth performance (median [IQR] $R^2$ of -0.009 [-0.014--0.0014]; two-sided Wilcoxon signed-rank $p=6.0\times10^{-8}$ for comparing LOOCV with RLOOCV; **Fig. 6c**).

Finally, we evaluated RLOOCV on real regression tasks, using datasets from UCIMLR (**Methods**). We used $L^2$-regularized regression, and employed nested 5-fold cross-validation to select the regularization parameters. As before, we also trained models that used only the first two principal components to emphasize the effects of distributional bias. We found a slight improvement with RLOOCV (median [IQR] $R^2$ of 0.15 [-0.01-0.40] for LOOCV with PCA vs. 0.17 [0.00-0.40] for RLOOCV with PCA, Wilcoxon signed-rank p=0.0024; and 0.44 [0.09-0.52] for LOOCV vs 0.44 [0.10-0.53] for RLOOCV, p=0.0039, respectively). While RLOOCV is less effective in regression compared to classification, these results demonstrate that distributional bias does manifest during LOOCV in regression, and can be alleviated with RLOOCV.

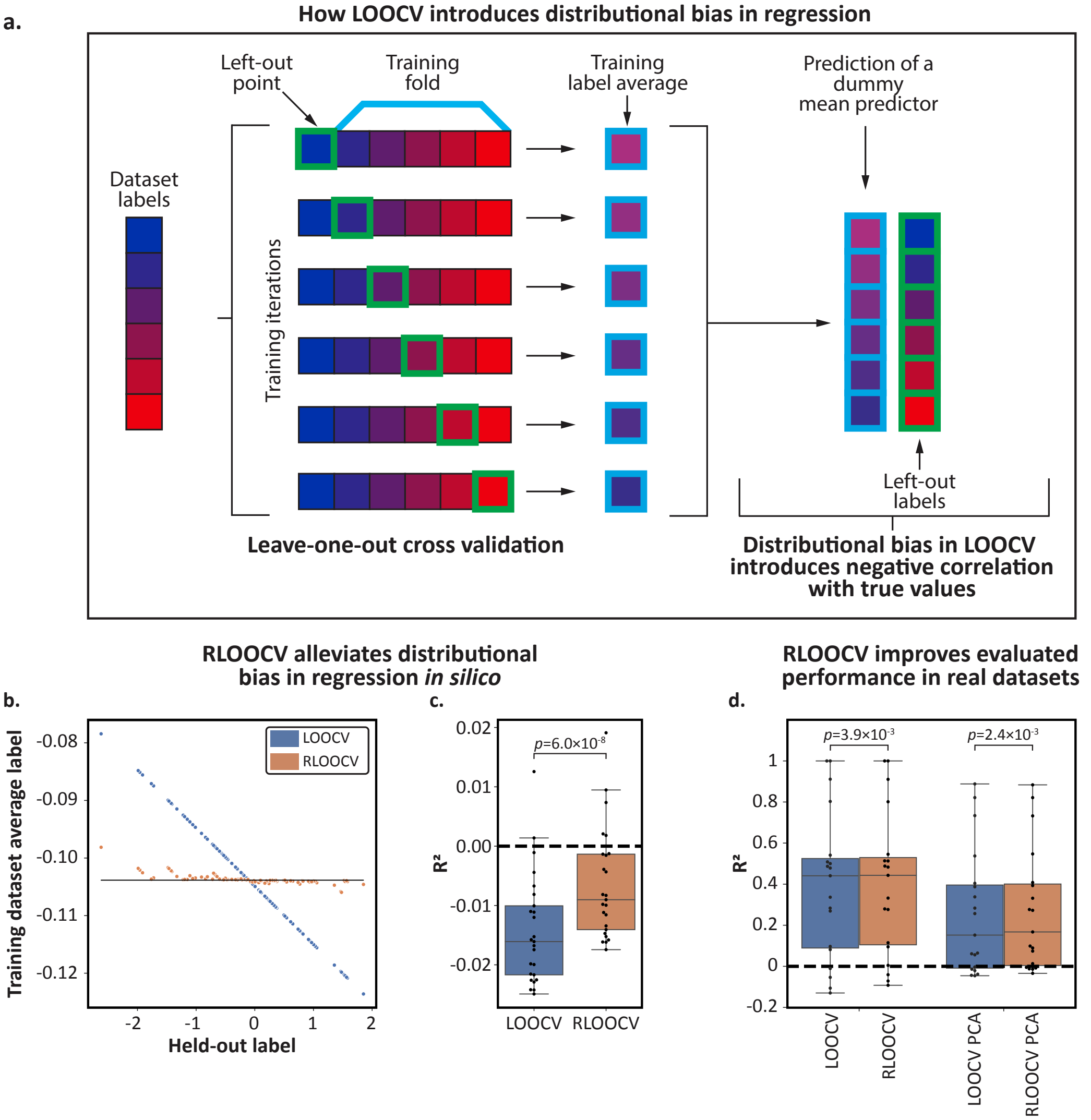

**Figure 6 | Distributional bias and RLOOCV generalize to regularization. a,** An illustration of how distributional bias occurs in LOOCV in regression. For any held-out instance, the average of the remaining dataset shifts slightly in the other direction. As a result, a dummy predictor that returns the average training class label would produce predictions that are perfectly inversely correlated with the held-out labels. **b,** An example of distributional bias manifesting in LOOCV of a simulated dataset (blue) and its absence in RLOOCV (orange). By selectively removing from the training dataset one additional data instance to shift the average as close as possible to the data average (black line), but not past it, we can alleviate the impact of distributional bias on the data. **c,** Synthetic simulations in which all data features and labels are random (**Methods**), meaning that a correct evaluation should yield $R^2$ of 0. LOOCV evaluations of $L^2$-regularized regression models yield median values less than 0 (one-sample t test vs. 0 $p=7.1\times10^{-8}$), while evaluations of RLOOCV demonstrate performances closer to the expected ground truth (Wilcoxon signed-rank $p=6.0\times10^{-8}$ comparing LOOCV with RLOOCV; one-sample t test vs. 0 $p=0.0017$). **d**, Evaluations of LOOCV and RLOOCV on regression tasks from UCIMLR (**Methods**), either considering all features or just the first two principal components. RLOOCV significantly outperforms LOOCV (Wilcoxon signed-rank $p=0.0039$; $p=0.0024$ for models run on the first two principal components).

# Discussion

Distributional bias emerges when there is a shift across the mean of the labels of different training sets, and predictions are evaluated jointly across all folds. We show that with an adversarial model, such as a dummy predictor that always predicts the negative of the average training-set label, this would result in information leakage and a perfect auROC and auPR, regardless of the underlying data. In practice, however, we show that since many machine learning models tend to regress to the mean, distributional bias would actually cause an under-evaluation of their predictive performance under LPOCV, especially when P is small or when the class ratio is imbalanced. We demonstrate this effect using multiple simulated and real datasets, and show that this phenomenon may also lead to a suboptimal selection of hyperparameters. To address this, we propose RLOOCV, a generalizable solution to maintain class balances across all training sets, by subsampling the training sets to ensure consistent label distributions.

While we demonstrate that distributional bias causes a consistent under-evaluation of performance using LOOCV, we note that the strength of this effect varies across models. In particular, it is strongest when models are highly regularized, in which case they are more likely to produce class probabilities that are closer to the training-set label average. In hyperparameter optimization, the distributional bias we observe in LOOCV would push towards selection of weaker regularization, which might then negatively affect subsequent generalization and performance on these models on additional datasets and limit interpretation and explainability of optimized models.

RLOOCV removes an additional training sample from the training set of each fold as a generalizable solution to distributional bias. We note, however, that there are a few other possible corrections that would address distributional bias: 1) stratified LPOCV (or stratified K-fold CV), so long that the class balance can be strictly maintained in all folds; 2) an "up-balanced" LPOCV, in which a sample from the same class is generated for each sample that is held out as a test set; 3) a modification of RLOOCV, in which for every left-out fold, a separate model is trained for every possible subsampled training dataset that maintains the desired class balance; and 4) similarly to (3) one could perform bagging[46] to train a collection of different models across different random subsets of the training set, or use other means for stratification, for example via stratified batch gradient descent, with each individual bag (or batch) from every training fold containing the same class balance. While, in general, it is possible to avoid distributional bias by subsampling training sets to smaller sizes and randomly selecting held-out test instances, implementing this contradicts many of the advantages of LOOCV, such as maintaining the largest sample set possible for training. Lastly, while we note that a post-hoc normalization of a model's predictions with respect to either the average training set labels or the model's predictions on the training set seems like an intuitive solution, we found scenarios in which this approach overestimated performance evaluation, such as K-Nearest Neighbors models with small K (**Supplementary Note 2**). Therefore, we do not recommend post-hoc normalization as a valid solution to distributional bias.

# Methods

## Synthetic simulations

Our simulation structure follows the standard approach across various fields, in which an entire dataset is first generated, and is then divided into different folds during cross-validation[18,28,31]. In this schem, datasets are i.i.d to one another, yet the folds of a given dataset are not i.i.d.[19,23]. This is similar to real-world applications, in which a dataset is generated once, and then divided into mutually-exclusive folds for cross-validation.

For each simulation, we randomly generated a set of N binary labels $y_i \in \{0,1\}$, for $i \in [1,2,\ldots,N]$, with class balances ranging from 0.1 to 0.9 in increments of 0.1. Data was generated by drawing for every sample a collection of 20 i.i.d features from a uniform distribution on [0,1]. The "negative-mean predictor" was defined as $\hat{y}_i = -\frac{\sum_{j \neq i} y_j}{N-1}$, where $\hat{y}_i$ is the prediction for each held-out sample $i$, $y_j \in \{0,1\}$ is the label for each sample j, with a total of $N$ samples across the entire dataset. For each simulation setting (e.g., each cell in **Fig. 2c**), we generated 100 random datasets of 250 or more data instances, where in certain scenarios we used more than 250 so that the P-left-out would divide evenly into the dataset size. In the case of random forest simulations (**Fig. S3a**), we simulated only 100 data instances per dataset to minimize computational runtime. For the Honest Leave-one-out and nested RLOOCV simulations (**Fig. S5a**), due to the high computational cost of running the nested tuning scheme, we evaluated the simulations using synthetic datasets of 50 samples. For the regression simulations (**Fig. 6c**), we generated the labels and features from normal distributions.

## Benchmarking of predictive analyses

We considered datasets which had previously been analyzed using LOOCV, had publicly available processed data, and had public code or clear methods. To this end, we considered datasets from Vogl et al[42], Lozano et al.[41], and Fettweis et al.[39]. The first two processed datasets and corresponding code were obtained from the original publications, while the third processed dataset was obtained from a different study with publicly available materials[40], which we processed as described, using a pseudocount of $10^{-6}$ and centered log ratio transform, followed by LOOCV and RLOOCV comparisons of tuned logistic regressors. In all cases, we show results from 10 bootstrap runs (**Fig. 4**). For the analysis of Lozano et al., we also demonstrate performance results across thirteen regularization strengths (**Fig. 5c**).

For the classification analyses of the datasets from the UC Irvine Machine Learning Repository[38], we accessed the data from the python ucimlrepo package. For the classification evaluations, we tested all the available datasets that included 'Classification' in their metadata. We subset each dataset to only the two most common target classes (to emulate binary classification), and then subsampled each dataset to only 50 randomly selected data instances for computational efficiency and to focus on the data-scarce scenarios in which LOOCV is typically used. We additionally required that a dataset included at least two distinct `y` classes in the first `target` column, with a minimum of two labels with over 5 counts per label (to support nested cross-validation), and at least two numeric features (X columns). This resulted in 49 datasets considered for analysis.

For the analyses of regression datasets, we considered all datasets labelled as "Regression", or included "Regression" and no "Classification" (as these datasets' labels were often better suited for classification evaluations rather than regression). For these tasks, we observed an over representation of labels on the max or min of different datasets, which appear to be samples that originally had missing values. We therefore filtered out all data instances whose y labels corresponded to the minimal or maximal value of their dataset. As before, we also randomly subsampled down to 50 data instances and required a minimum of two numeric feature columns. Both the labels and the features were standard scaled. To ensure we were utilizing datasets with labels that were reasonable to model using linear regression, we only considered the datasets whose `y` labels included a minimum of 10 distinct values, which resulted in 19 datasets. To address the smaller number of datasets and potential noise introduced during subsampling, we repeated this procedure 5 times for each regression dataset, and evaluated each method's performance by the mean $R^2$ across all 5 runs. For all evaluations, we compared performances of $L^2$-regularized regression, considering either the entire dataset or its first two principal components. We trained and evaluated these models using both LOOCV and RLOOCV, and tuned scikit-learn's regularization hyperparameters, within an inner (nested) 5-fold CV, Honest LOOCV, or nested RLOOCV (**Fig. 6d, Fig. S5**).

**Code availability**

All code from this analysis is publicly available at https://github.com/korem-lab/LPOCV-bias-analysis. An easy-to-use package to facilitate use of rebalanced cross-validation approaches can be found at https://github.com/korem-lab/RebalancedCV.

**Supplementary Note 1 | Generalizing RLOOCV to RLPOCV and RKFold**

Consider a dataset consisting of $T$ positive samples, $F$ negative samples, and a desired $P$ left-out schema. The simple approach to rebalance LPOCV per held-out set is to randomly remove $P$ cases from the training set with labels opposite those of the held-out set, such that every model is trained using exactly $T + F - 2P$ observations, consisting of $T - P$ positive and $F - P$ negative samples. However, acknowledging that LPOCV would be considered in scenarios that are already data-sparse, we show that using stratification for the held-out test set, it is possible to develop a stratified RKFold schema that removes fewer samples from the training set, which can be especially important in a highly imbalanced dataset with few observations from one class. We consider the following:

- $N = T + F//P$, the total number of $P$-sized held-out groups.

- $T_c = T//N$, the total number of positive class observations that are guaranteed to be present in every left-out $P$ group.
- $F_c = F//N$, the total number of negative class observations that are guaranteed to be present in every left-out $P$ group.

Therefore, every held-out group consists of:

1. $T_c$ positive samples.
2. $F_c$ negative samples.
3. $P - (T_c + F_c)$ remaining samples, which will cause label averages to vary across training folds, leading to distributional bias.

Because components (1) and (2) are constant across all training folds, there is no need to subsample training datasets to match those held out samples. The only subsampling needed to ensure consistent class balances across all training folds is to remove samples opposite each case in (3). Therefore, when performing stratified RLPOCV, each training fold should use exactly:

- $T - P + F_c$ positive samples
- $F - P + T_c$ negative samples

To demonstrate a few examples, under this RLPOCV implementation, for a dataset with T=50, F=50, and P=5, each model will be trained using exactly 47 T and 47 F samples. For a different dataset with T=10, F=1000, and P=2, each model will be trained using exactly 9T and 998 F samples. We note that in the case of P=1, RLPOCV is identical to the RLOOCV implementation above.

**<u>Supplementary Note 2 | Post-hoc standardization addresses distributional bias but may lead to performance inflation.</u>**

An intuitive approach for correcting distributional bias could be to normalize (scale, z-score, etc.) a model's predictions on the held-out test set with respect to the same model's predictions on its training set. We describe this approach here as a cautionary tale, as it is unsuitable in certain settings. In some of our simulations, such as those analyzed with logistic regression, this approach performs exactly as intended, resulting in auROCs close to the expected 0.5 ($p = 0.14$ via 1-sample t-test; **Fig. S7a**). However, in cases in which a model does not have the flexibility to precisely predict a class balance, for instance only being capable of predicting 0 or 1, then any normalization with respect to training label averages will create perfect rankings within all samples initially predicted at the same value. The simplest example of this is the K-Nearest Neighbor model with K=1: When running this predictor through our same simulation framework but including a post-hoc z-score standardization to the training predictions, we observe auROCs larger than the intended 0.5 in every case ($p < 0.001$ via 1-sample t-test; **Fig. S7b**), and with auROCs greater than 0.8 at extreme class imbalances. Although there are cases in which this normalization strategy is not problematic, we see no generalizable and concrete methods to ensure that elements of the more problematic models are not present in a given situation. Therefore, we conclude that there is always some risk of over-evaluation of machine learning performance when using a post-hoc normalization, and we recommend to avoid using it as a solution to distributional bias. Instead, RLOOCV solves this challenge without introducing any risk of overinflating results.

## Supplementary Figures

a.

Standard cross-validation in machine learning

Validation fold
Training fold
Validation predictions
Evaluate predictions
Combine evaluations
Training iterations
Performance$_1$
Performance$_2$
Performance$_3$
Performance$_4$
Performance$_5$
Performances

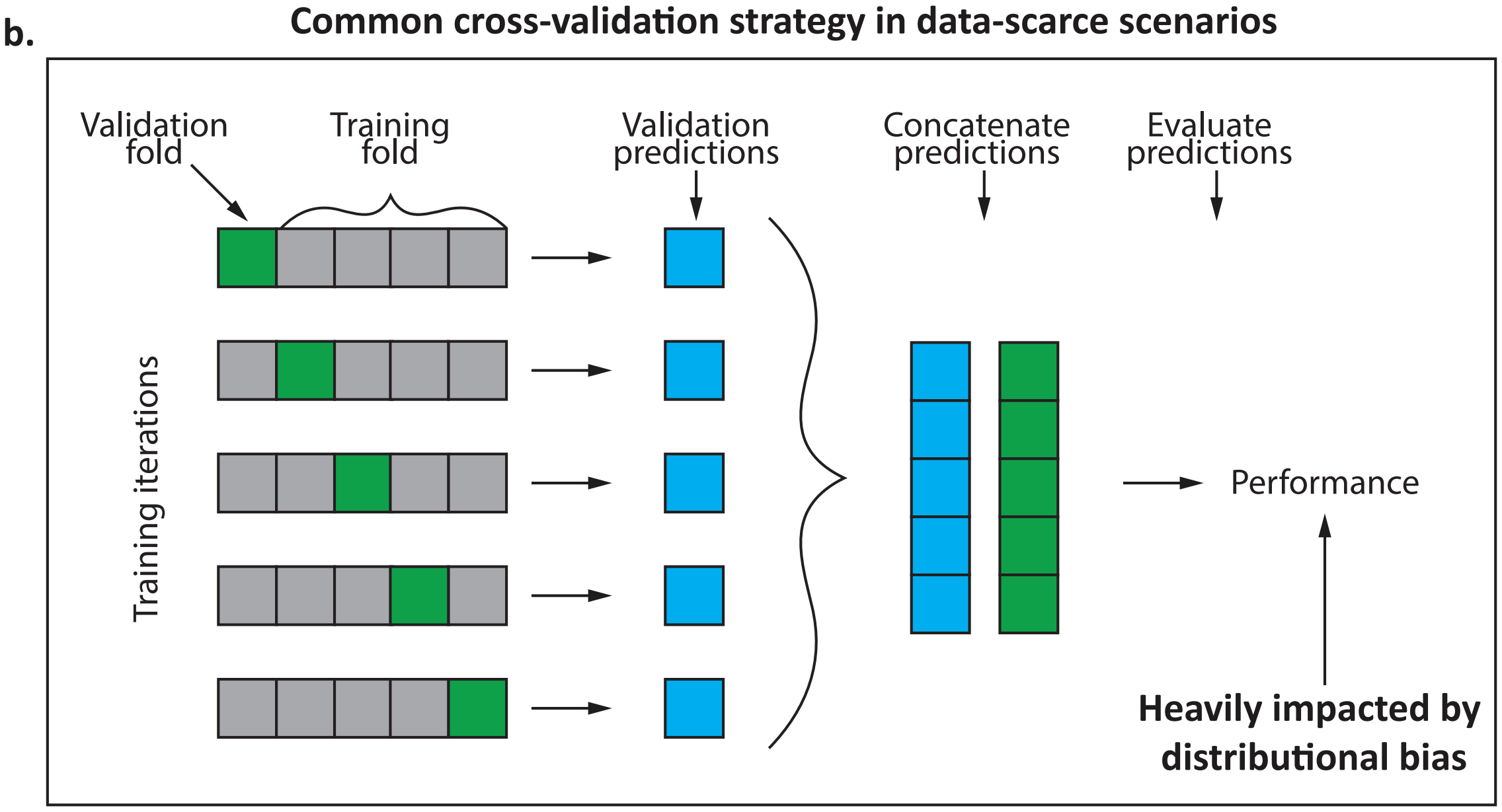

**Figure S1 | Comparison of cross-validation strategies. a,** Description of standard cross-validation approaches in machine learning, in which performance evaluation is performed separately for each held-out test set. **b,** A commonly used cross-validation strategy in data-scarce scenarios, in which the evaluation across all folds is performed once on the concatenation of all predictions. The evaluation approach in (**b**) can be heavily impacted by distributional bias. Figure inspired by ref. [6].

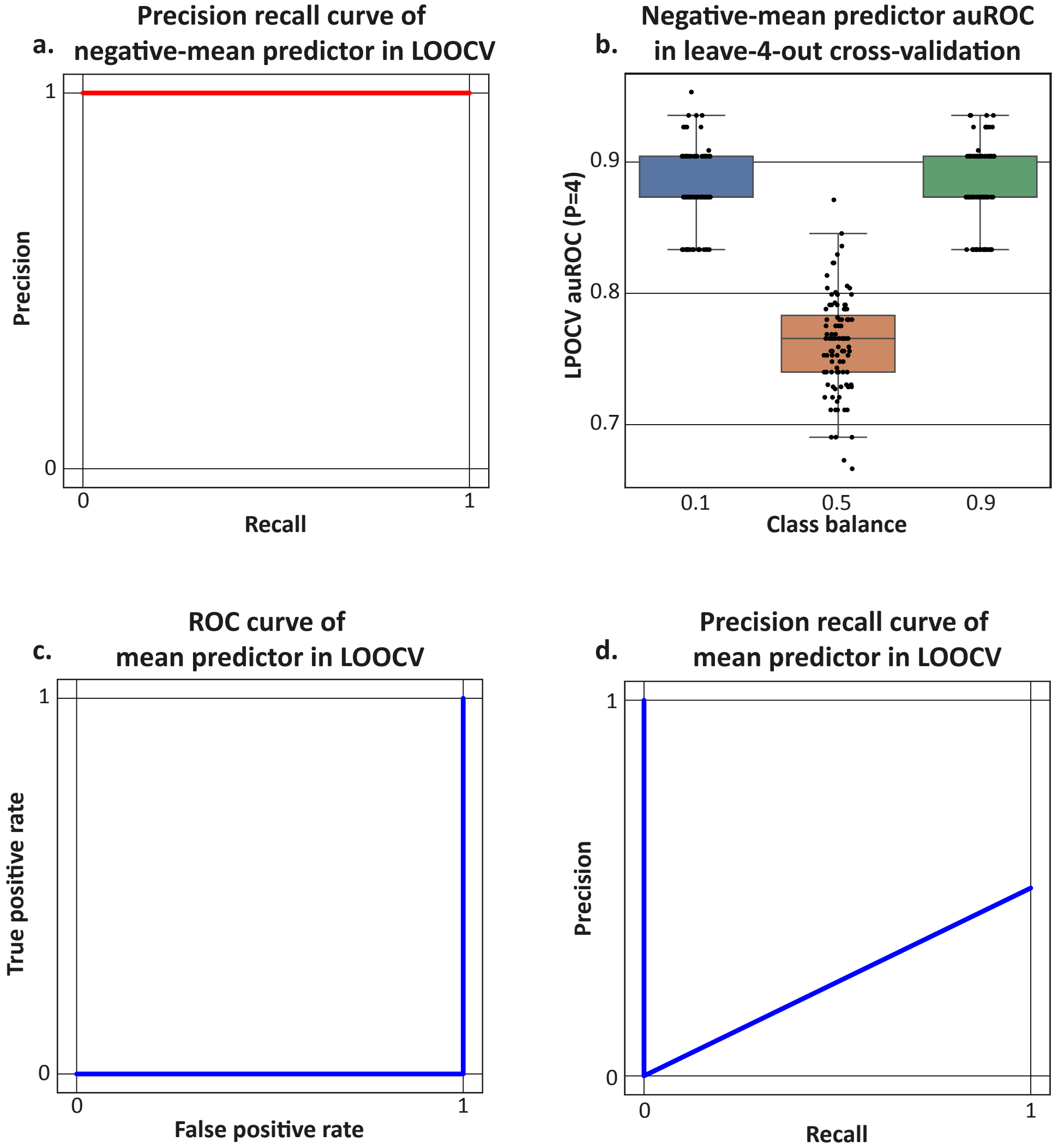

**Figure S2 | Impact of distributional bias across dummy models and evaluation metrics. a,** Results from the same model as in **Fig. 1b**, shown in a precision-recall curve, with an area under the curve of 1. **b,** Box and swarm plot demonstrating results from the same set of models as in **Fig. 1c** for P=4 and class balances of 10%, 50%, and 90%. Box, IQR; line, median; whiskers, nearest point to 1.5*IQR. **c,d,** ROC and PR curves similar to **Fig. 1b** and **Fig. S2a**, but for a predictor that outputs the label mean of the training set. Under LOOCV evaluation, for a class balance of 0.5, auROC=0 and auPR=0.25.

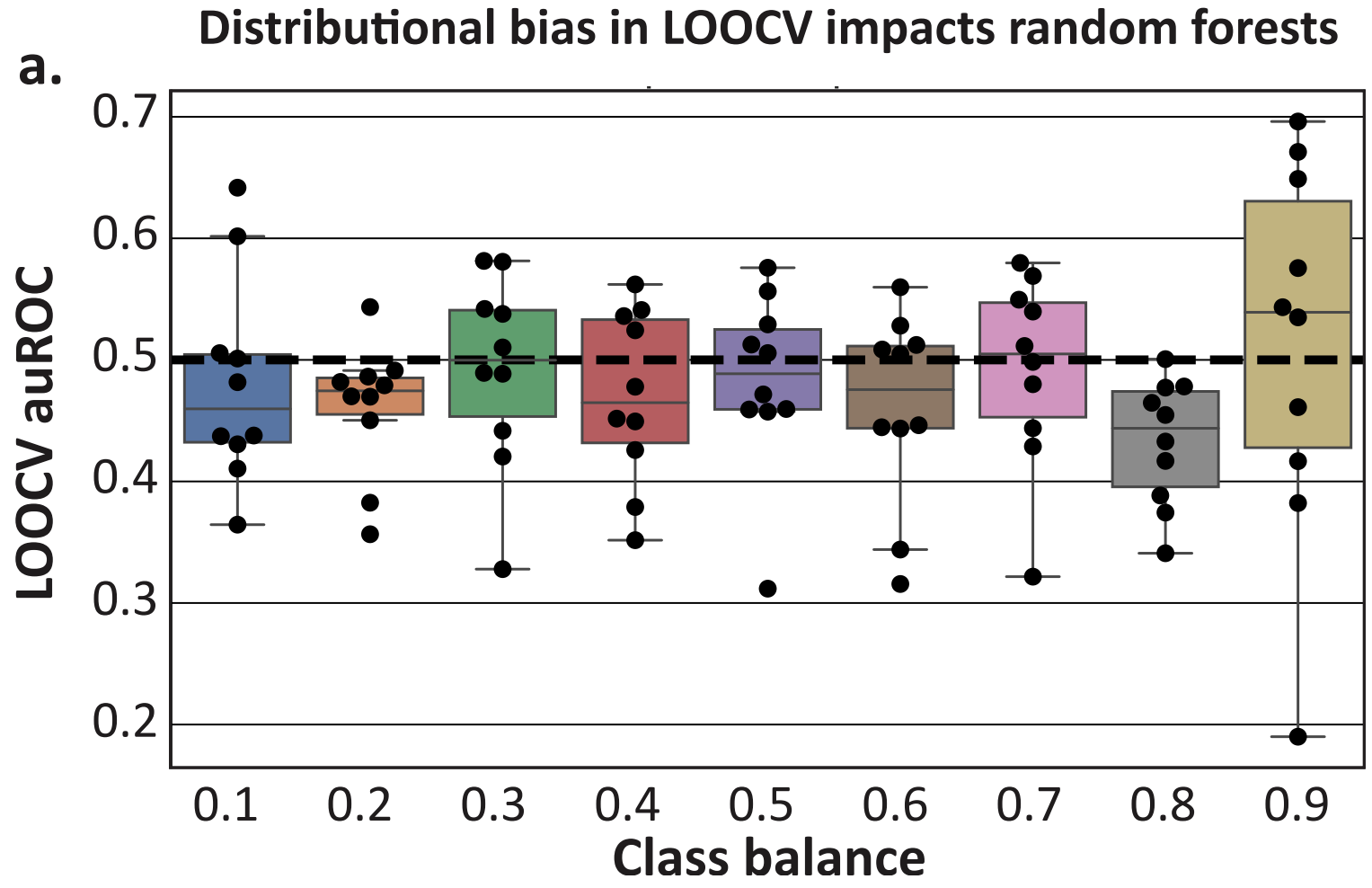

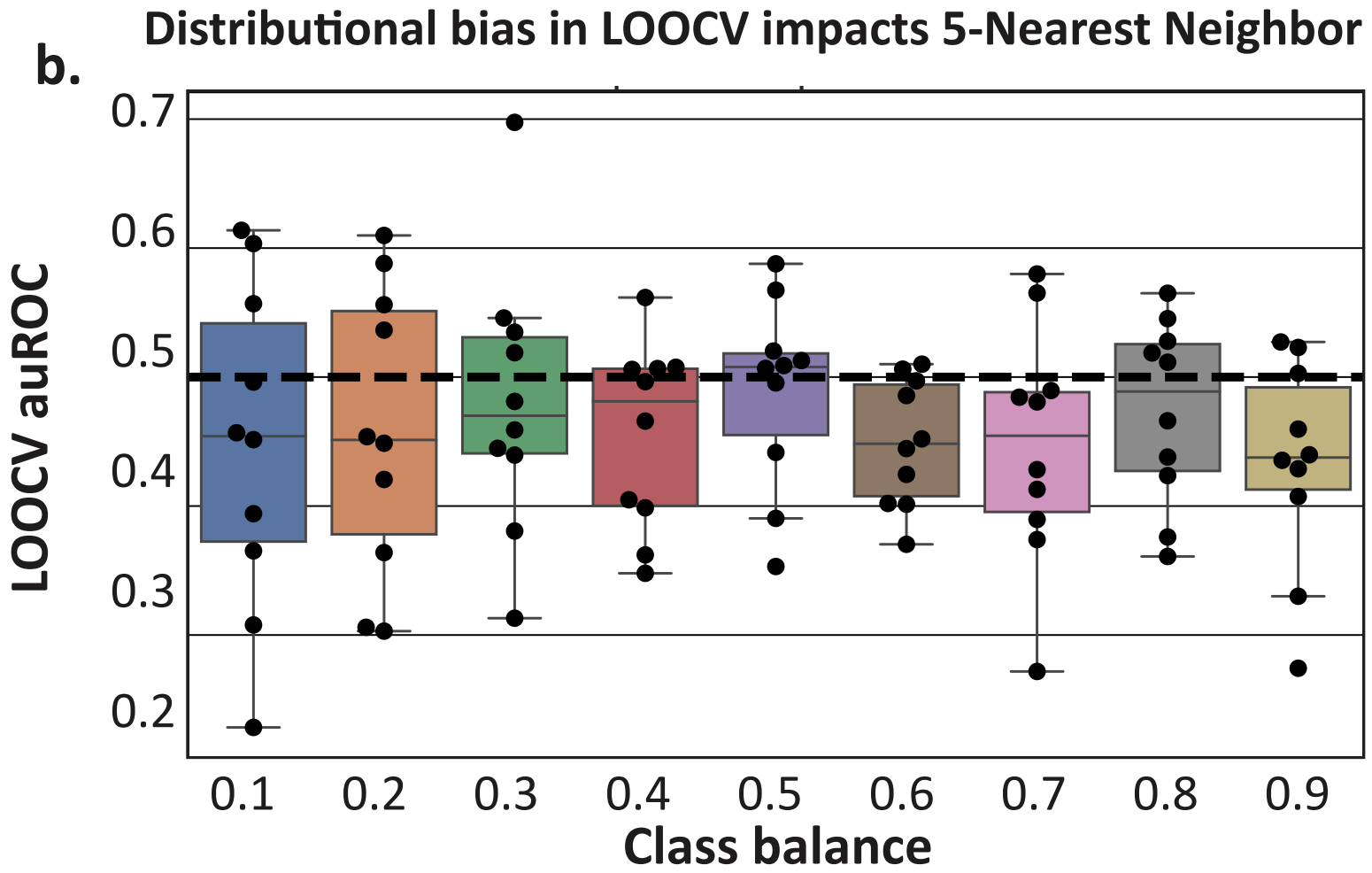

**Figure S3 | Distributional bias impacts random forest and 5-Nearest Neighbor models. a,b,** Boxplots demonstrating similar analyses as in **Fig. 2a**, but for random forest models (a), and K-Nearest Neighbors with K=5 (b). The resulting auROCs are consistently lower than 0.5 ($p$=0.009 via one-sample t-test vs. 0.5 for random forest, $p=2.9\times10^{-5}$ for K-Nearest Neighbors). Box, IQR; line, median; whiskers, nearest point to 1.5*IQR.

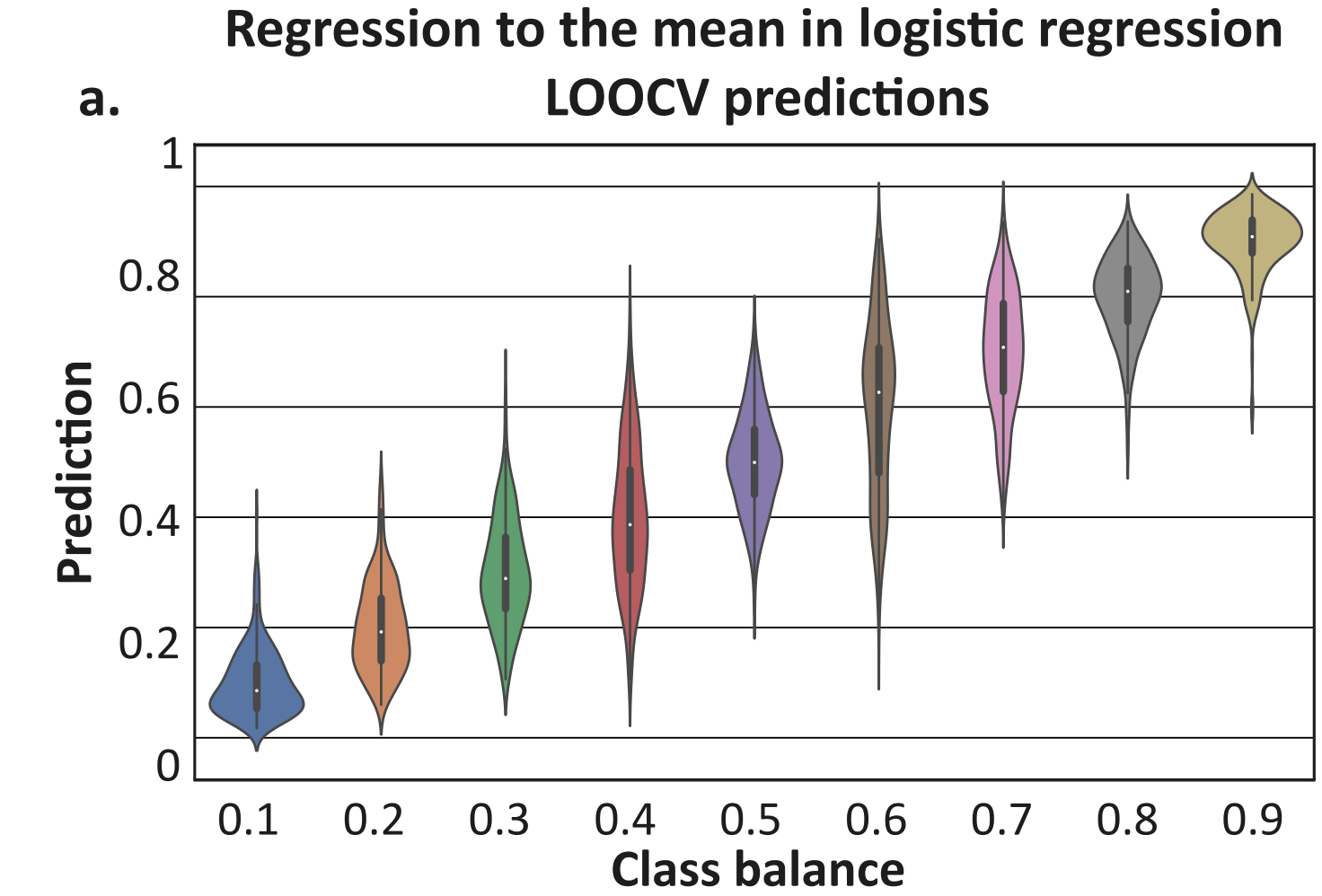

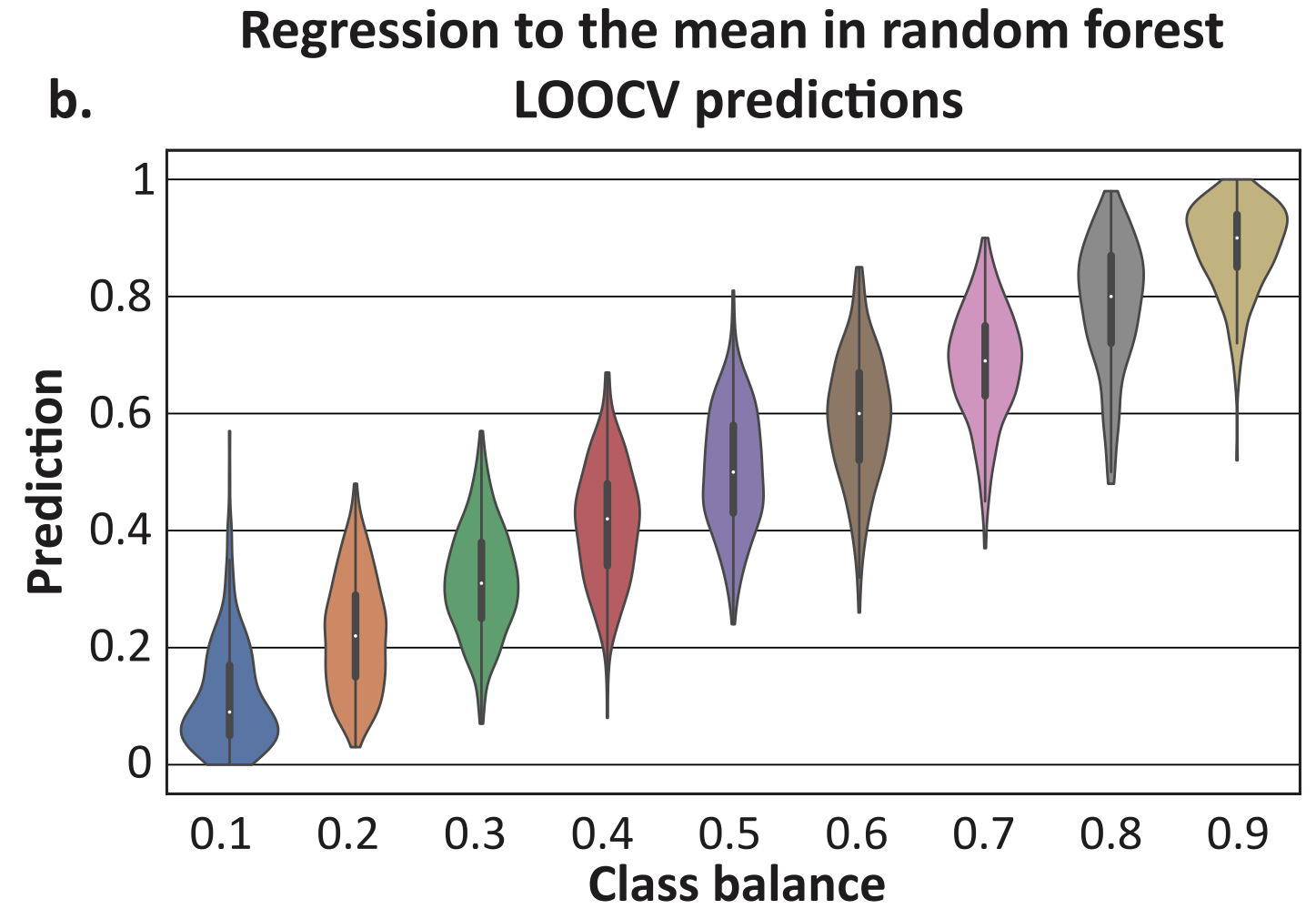

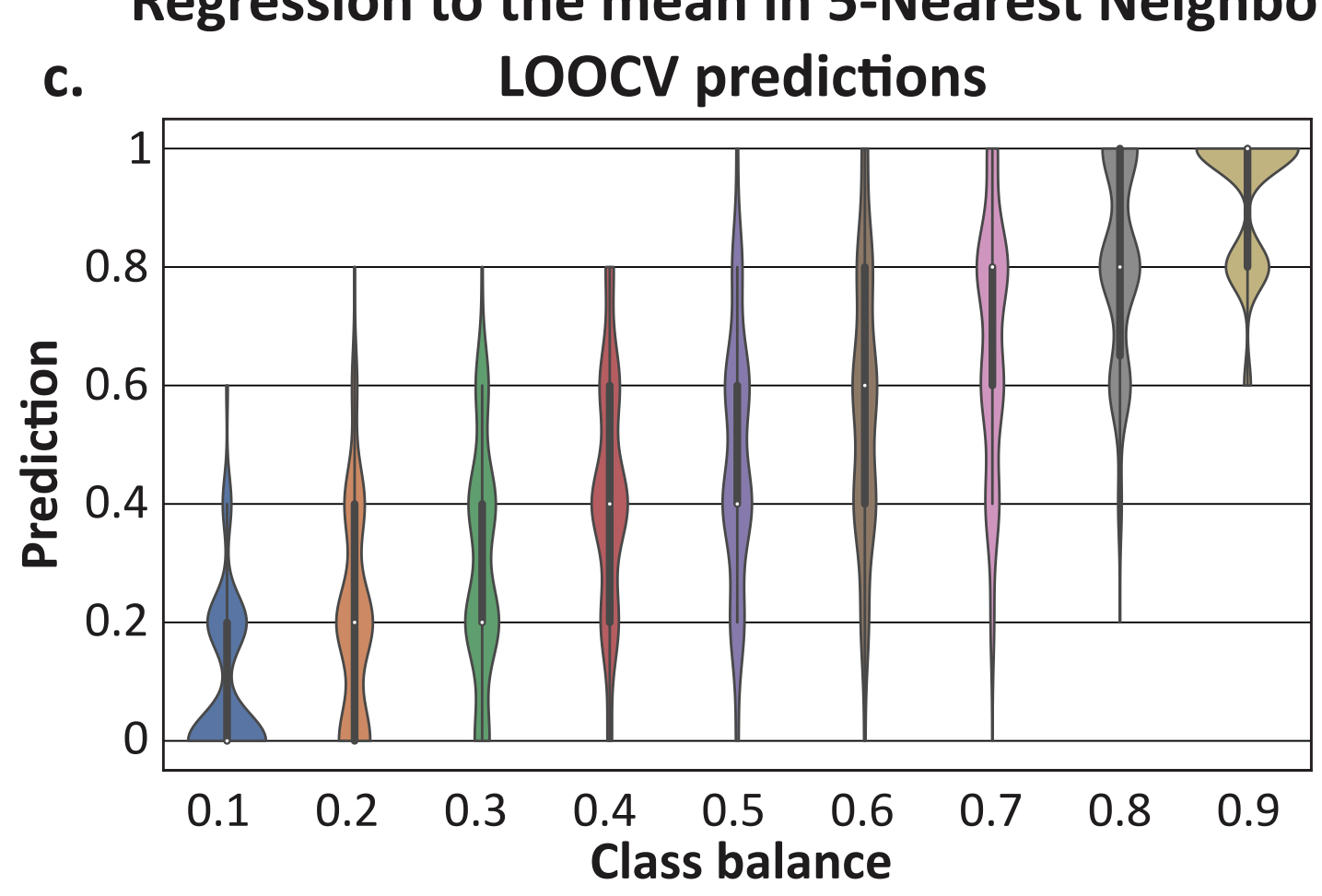

**Figure S4 | Regression to the mean in common machine learning models. a-c,** Violin plots showing predictions of logistic regression (**a**), random forest (**b**), and 5-Nearest Neighbors (**c**) models on randomly generated datasets in LOOCV, across different underlying class balances. All models used default scikit-learn parameters.

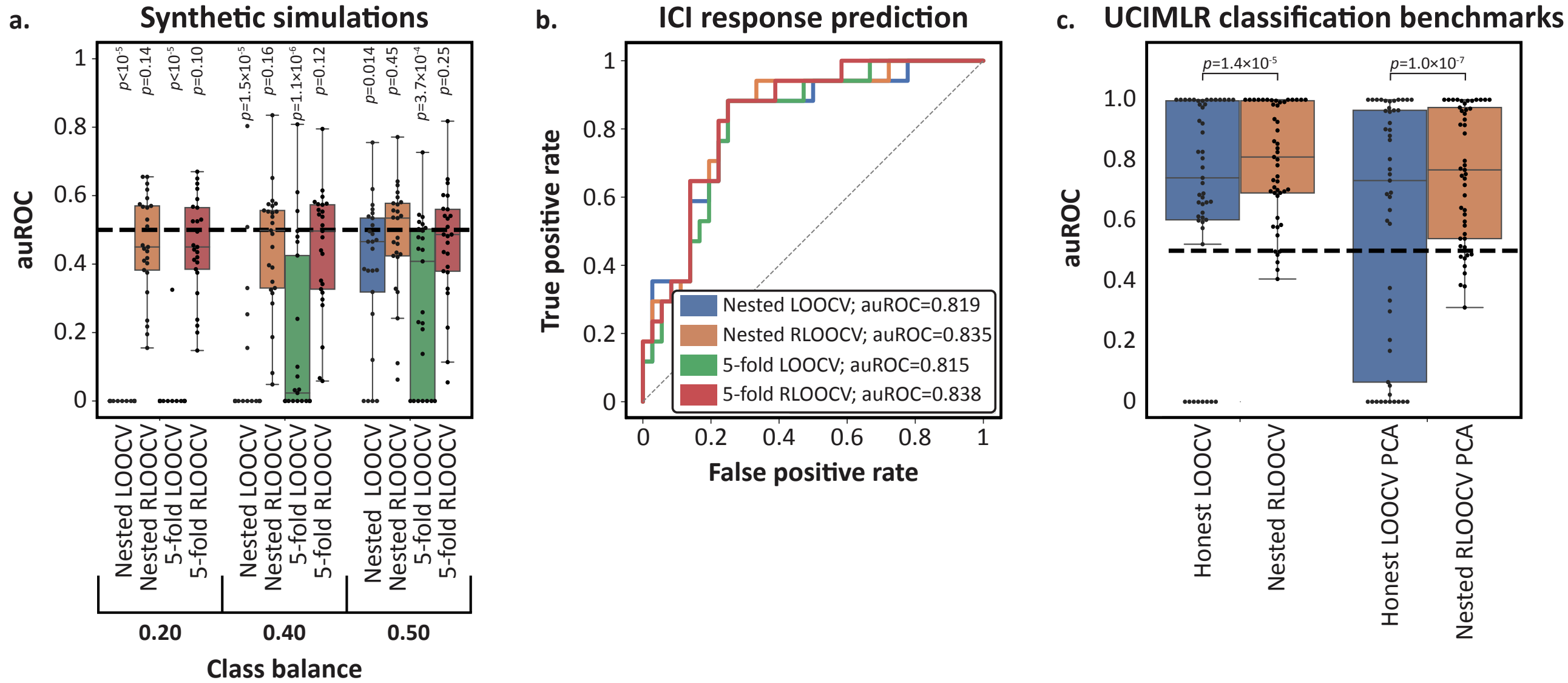

**Figure S5 | Distributional bias compromises nested cross-validation and Honest Leave-one-out.** All analyses pertain to comparison of hyperparameter optimization schemes using nested cross-validation, including either: (1) outer LOOCV and inner LOOCV (known as "Honest LOOCV"[31]); (2) outer and inner RLOOCV ("Nested RLOOCV"); (3) outer LOOCV with inner 5-fold cross validation ("5-fold LOOCV"); and (4) outer RLOOCV with inner 5-fold cross-validation ("5-fold RLOOCV"). In all cases, the nested cross-validation was used to tune the $L^2$ regularization of logistic regression models. **a,** Synthetic simulations on randomly generated data (as in **Figs. 2,3**; **Methods**). **b**, ROC curve of models trained on the classification task from Lozano et al.[41] (as in **Fig. 5c**). **c**, Results on classification tasks from data published in the UCIMLR, as in **Fig. 4a**, for logistic regression models evaluated either with Honest LOOCV (i.e. nested LOOCV) or nested RLOOCV.

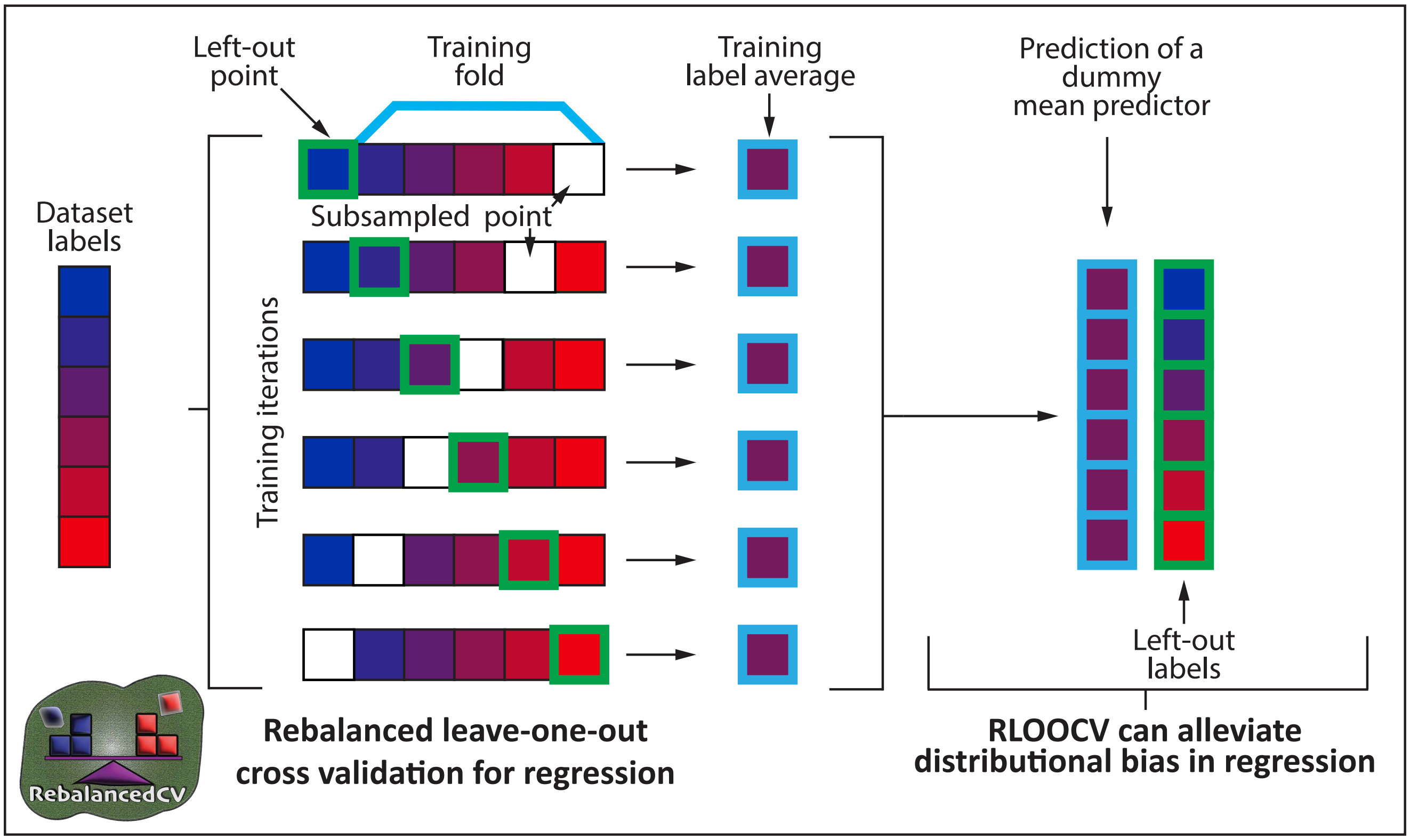

**Figure S6 | RLOOCV can alleviate distributional bias in regression.** An illustration of our proposed rebalanced LOOCV (RLOOCV) approach for regression. For each test instance (or fold), we remove from the training set a data instance that is as close to the opposite of the held-out test-set difference to the dataset mean, while not surpassing this value (i.e. the supremum). Doing this alleviates the impact of distributional bias on the dataset without risking an overcorrection that could erroneously inflate the evaluated performance.

**Standardizing test predictions with respect to a model's training set predictions is not a viable distributional bias correction method**

**a.** **Standardizing auROCs with Logistic Regression on random data**

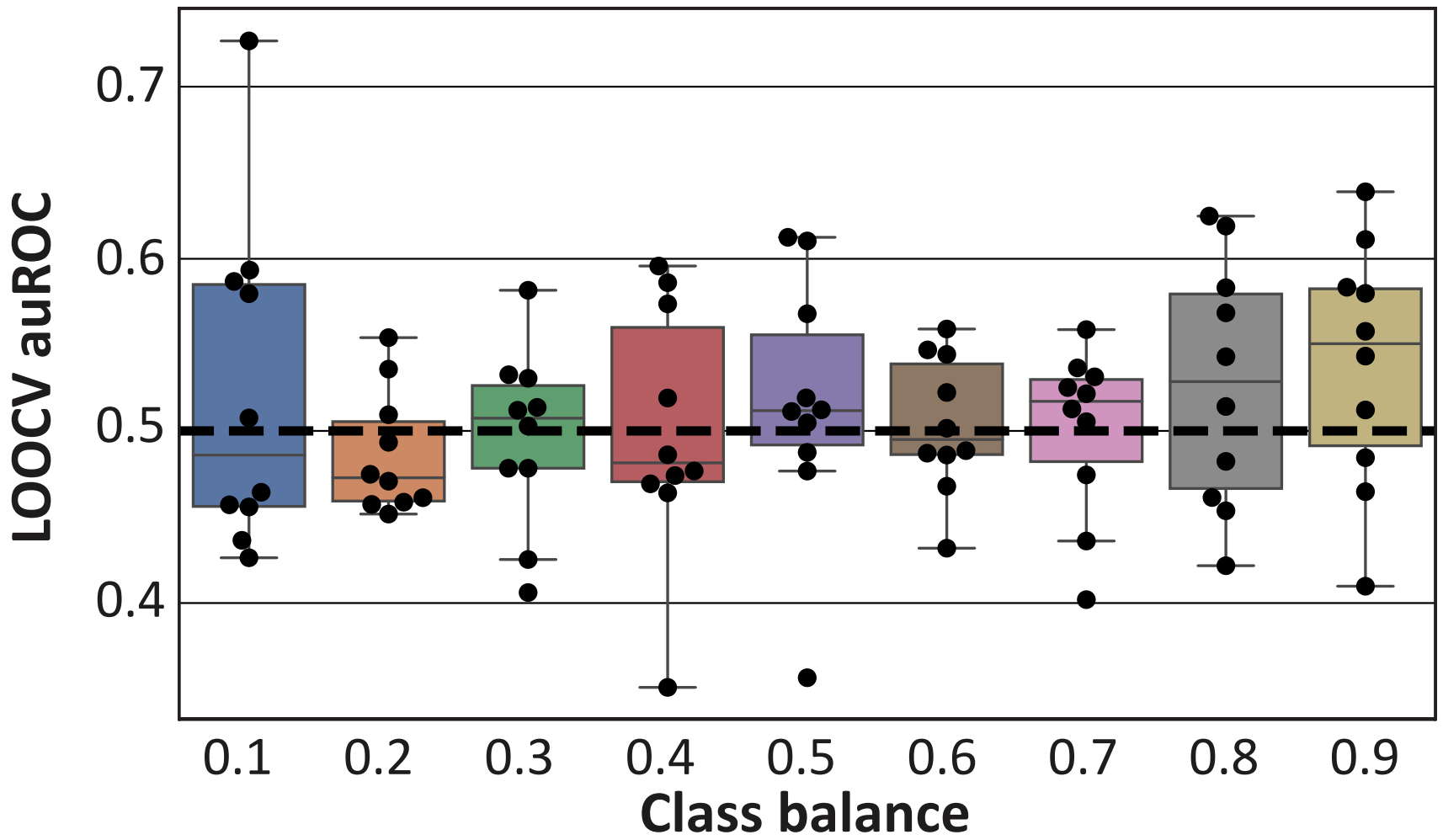

**b.** **Standardizing auROCs with Nearest Neighbor on random data**

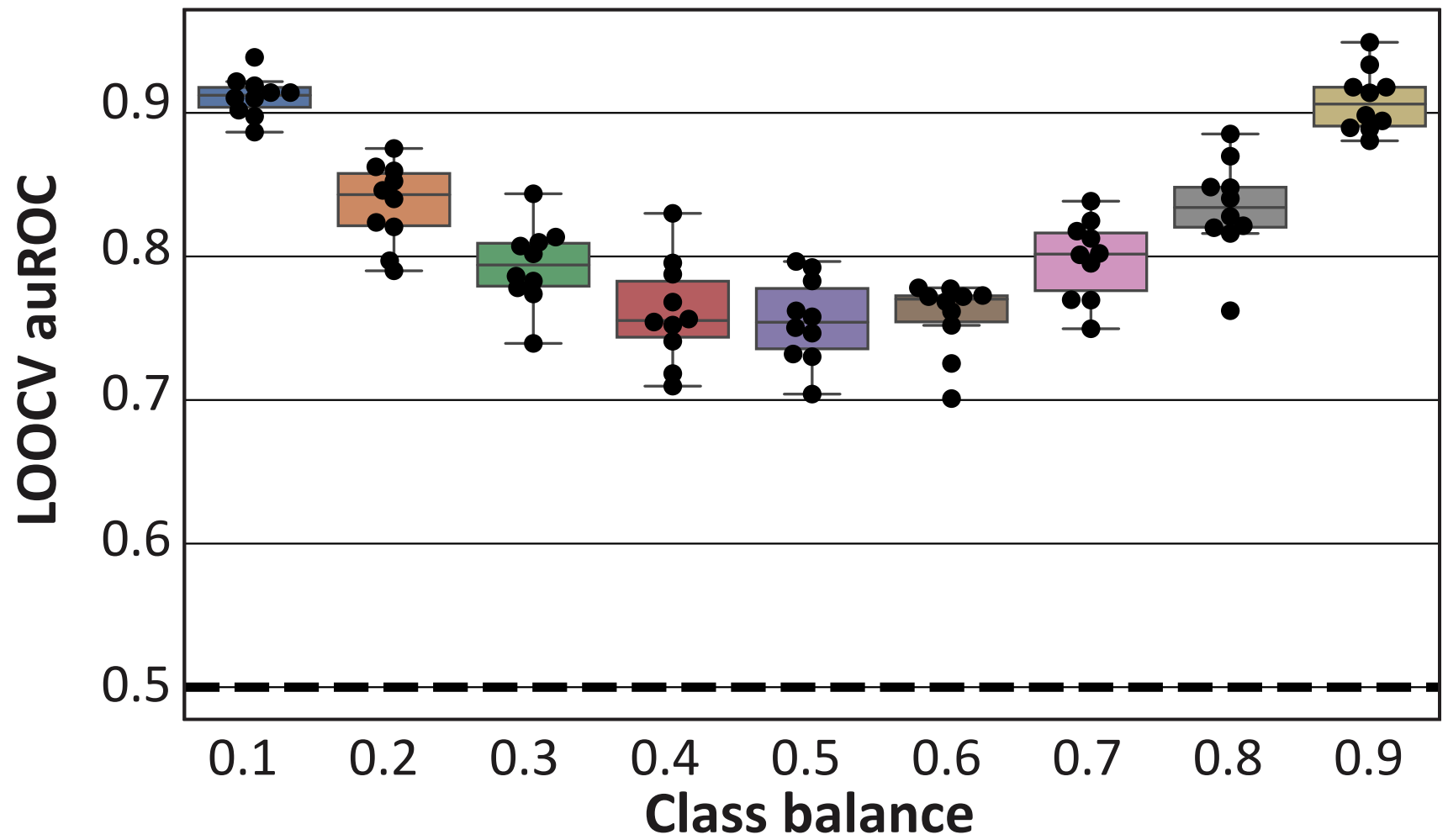

**Figure S7 | Post-hoc prediction standardization may lead to over-evaluation of performance.** Boxplots demonstrating a similar analysis as in **Fig. 2a**, but all test predictions are standardized to the model's predictions on the training set via z-scoring. **a,** Logistic regression evaluation standardized in this way produces results similar to the expected random guess, with an auROC close to 0.5 ($p = 0.14$ via a single 1-sample t-test of the aggregated results). **b,** 1-Nearest Neighbor models in the same standardized evaluation framework yield results larger than 0.5 ($p < 0.001$ via a single 1-sample t-test of the aggregated results). Box, IQR; line, median; whiskers, nearest point to 1.5*IQR.